%% file: main.tex
\documentclass[a4paper,11pt]{article}
\pdfoutput=1 

\input cyracc.def

\usepackage{jheppub} 
\usepackage{shuffle}

\usepackage[T1]{fontenc} 

\usepackage{lmodern}

\usepackage{bbm}
\usepackage{xcolor}
\usepackage{mathtools}
\usepackage{mathdots}
\usepackage{amsmath}
\usepackage{bm}
\usepackage{amsfonts}
\usepackage{dsfont}
\usepackage{amsthm}
\usepackage{enumitem}
\usepackage{booktabs}
\usepackage{siunitx}
\usepackage{arydshln}

 \usepackage{tikz}
    \usetikzlibrary{positioning}
    \usetikzlibrary{arrows}
    \usetikzlibrary{shapes}
    \usepackage{pgfplots}
    \usetikzlibrary{calc}
    \usetikzlibrary{decorations.markings}
     \usetikzlibrary{decorations.pathmorphing}
    \tikzset{snake it/.style={decorate, decoration=snake}}
\def\centerarc[#1](#2)(#3:#4:#5) 
    { \draw[#1] ($(#2)+({#5*cos(#3)},{#5*sin(#3)})$) arc (#3:#4:#5); }
    \usepackage{booktabs}
\usepackage{multirow,ctable,colortbl}

\definecolor{cnblau}{RGB}{0,80,147}

\newcolumntype{?}{!{\vrule width 1.5pt}}
\newenvironment{gleichung*}{\begin{equation*}\begin{aligned}}{\end{aligned}\end{equation*}}

\newcommand{\cL}[0]{\mathcal{L}}

\newcommand{\ord}[0]{\mathcal{O}}
\newcommand{\cI}[0]{\mathcal{I}}
\def\beq{\begin{equation}}
\def\eeq{\end{equation}}
\def\bsp#1\esp{\begin{split}#1\end{split}}

\newcommand{\ux}{\underline{x}}

\newcommand{\rd}{\textrm{d}}
\newcommand{\ut}{\underline{t}}

\newcommand{\xh}{{x^\prime}}
\newcommand{\bfW}{{\bf W}}

\title{The ice cone family and iterated integrals for Calabi-Yau varieties}

\author[a]{Claude Duhr}
\author[ab]{Albrecht Klemm}
\author[c]{Christoph Nega}
\author[c]{Lorenzo Tancredi}

\affiliation[a]{Bethe Center for Theoretical Physics, Universit\"at Bonn, D-53115, Germany}
\affiliation[b]{Hausdorff Center for Mathematics, Universit\"at Bonn, D-53115, Germany}
\affiliation[c]{Physics Department, Technical University of Munich, D-85748 Garching, Germany}

\emailAdd{cduhr@uni-bonn.de}
\emailAdd{aklemm@th.physik.uni-bonn.de}
\emailAdd{c.nega@tum.de} 
\emailAdd{lorenzo.tancredi@tum.de}

\abstract{
We present for the first time fully analytic results for multi-loop equal-mass ice cone graphs in two dimensions. By analysing the leading singularities of these integrals, we find that the maximal cuts in two dimensions can be organised into two copies of the same periods that describe the Calabi-Yau varieties for the equal-mass banana integrals. We obtain a conjectural basis of master integrals at an arbitrary number of loops, and we solve the system of differential equations satisfied by the master integrals in terms of the same class of iterated integrals that have appeared earlier in the context of equal-mass banana integrals. We then go on and show that, when expressed in terms of the canonical coordinate on the moduli space, our results can naturally be written as iterated integrals involving the geometrical invariants of the Calabi-Yau varieties. Our results indicate how the concept of pure functions and transcendental weight can be extended to the case of Calabi-Yau varieties. Finally, we also obtain a novel representation of the periods of the Calabi-Yau varieties in terms of the same class of iterated integrals, and we show that the well-known quadratic relations among the periods reduce to simple shuffle relations among these iterated integrals.
}

\begin{document}
\preprint{\begin{minipage}[t]{8cm}\begin{flushright} 
BONN-TH-2022-24 \\
TUM-HEP-1444/22
      \end{flushright}\end{minipage}}

\maketitle

\flushbottom


\input{intro}

\input{icecone}

\input{structure}

\input{conclusions}

\section*{Acknowledgment}
CN would like to thank Kilian B\"onisch and Nikolaos Syrrakos for discussions and help with the numerics. CN and LT were supported by the Excellence
Cluster ORIGINS funded by the Deutsche Forschungsgemeinschaft (DFG, German Research
Foundation) under Germany's Excellence Strategy - EXC-2094 - 390783311 and by the ERC
Starting Grant 949279 HighPHun. AK and CN likes to thank Dr. Max R\"ossler, the Walter Haefner Foundation and the ETH Z\"urich Foundation for support.


\appendix

\input{appendices}

\input{shuffle}


\bibliographystyle{JHEP}
\bibliography{References}

\end{document}

%% file: intro.tex

\section{Introduction}
\label{sec:intro}

Multi-loop Feynman integrals are a cornerstone of perturbative Quantum Field Theory (QFT) and of many modern methods to make precise predictions for collider and gravitational wave experiments. Consequently, developing efficient techniques for their evaluation and a solid understanding of the mathematics underlying them, including the special functions they evaluate to, has been a major area of research since the early days of QFT. Unitarity implies that scattering amplitudes and Feynman integrals must have non-trivial discontinuities, and so it is generically expected that multivalued special functions show up in this context. Having a well-defined class of special functions allows one to study their properties in an independent fashion. This is often useful in applications, as it allows one, e.g., to make sure that the results have no `hidden zero', i.e., no subset of terms evaluates to zero. Already several decades ago, it became clear that the logarithm and the dilogarithm prominently show up in one-loop computations in four space-time dimensions~\cite{tHooft:1978jhc}. It was quickly realised that these functions are insufficient at higher loops, and also multiple polylogarithms (MPLs) appear~\cite{Kummer,Lappo:1927,GoncharovMixedTate,Goncharov:1998kja,Brown:2011ik,Remiddi:1999ew,Gehrmann:2000zt,Ablinger:2011te}, defined as iterated integrals involving rational kernels~\cite{ChenSymbol}. This class of functions is mathematically well understood (cf., e.g., refs.~\cite{Goncharov:1995,Vollinga:2004sn,Goncharov:2005sla,Brown:2008um,Goncharov:2009tja,Brown:2009qja,Goncharov:2010jf,Buehler:2011ev,Duhr:2011zq,Duhr:2012fh,Ablinger:2013cf,Duhr:2014woa,Frellesvig:2016ske,Ablinger:2018sat,Naterop:2019xaf,Duhr:2019tlz}).

It was noted as early as 1964 that not every Feynman integral can be expressed in terms of MPLs, but starting from two loops in four dimensions also elliptic integrals arise in the computation of the two-loop two-point function with three massive propagators (the so-called two-loop \emph{banana integral})~\cite{Sabry}. While over the following decades evidence was mounting that functions associated to elliptic curves play an important role in QFT~\cite{Broadhurst:1987ei,Bauberger:1994by,Bauberger:1994hx,Laporta:2004rb,Kniehl:2005bc,Aglietti:2007as,Brown:2013hda,CaronHuot:2012ab}, it took nearly half a century to identify~\cite{Bloch:2013tra} the relevant class of special functions as elliptic multiple polylogarithms (eMPLs)~\cite{MR1265553,LevinRacinet,BrownLevin,Broedel:2017kkb} and iterated integrals of modular forms~\cite{ManinModular,Brown:mmv}.\footnote{See also refs.~\cite{Adams:2016xah,Adams:2014vja,Adams:2013nia,Adams:2015gva,Ablinger:2017bjx} for closely related classes of special functions.} These functions have by now seen many applications in the context of Feynman integrals (see ref.~\cite{Bourjaily:2022bwx} for a recent review of these functions and their applications in physics). In particular, the two-loop banana integral can be expressed in terms of eMPLs~\cite{Bloch:2013tra,Broedel:2017siw,Campert:2020yur,Bogner:2019lfa}. When the masses are equal, the integral can also be written in terms of iterated integrals of modular forms, which are the natural iterated integrals associated with the moduli space of this family of elliptic curves~\cite{Adams:2017ejb,Broedel:2018iwv}.

While special functions associated with families of elliptic curves and their moduli spaces cover large ranges of interesting Feynman integrals, it is known that at higher loops also special functions associated to Calabi-Yau varieties arise~\cite{Brown:2010bw,Bloch:2014qca,Bloch:2016izu,Bourjaily:2018yfy,Bourjaily:2019hmc,Bourjaily:2018ycu,Vergu:2020uur,Duhr:2022pch}. The simplest representatives are higher-loop banana integrals, and the geometry attached to an $l$-loop banana integral is a Calabi-Yau $(l-1)$-fold~\cite{Bloch:2014qca,Bloch:2016izu,Fischbach:2018yiu,Klemm:2019dbm}. The associated special functions are solutions to inhomogeneous extensions of Picard-Fuchs differential systems. The corresponding homogeneous solutions describe the periods of the family of Calabi-Yau varieties, and a lot is known about them in the mathematics and string theory literature. It is then possible to use standard techniques to obtain series representations for the inhomogeneous solutions~\cite{Fischbach:2018yiu,Klemm:2019dbm}. In the special case of equal-mass banana integrals in $d=2$ dimensions, one can write the result in terms of an integral involving the periods of the Calabi-Yau variety~\cite{Bonisch:2021yfw}. Specialising to the three-loop case, the underlying geometry is a one-parameter family of K3 surfaces, and it is possible to express the result in terms of iterated integrals of (meromorphic) modular forms~\cite{Broedel:2019kmn,Broedel:2021zij,Pogel:2022yat}.\footnote{See also refs.~\cite{Primo:2017ipr,Bezuglov:2021jou,Kreimer:2022fxm} for other classes of iterated integrals that arise in the case of the three-loop equal-mass banana integrals, and ref.~\cite{Forum:2022lpz} for first steps towards understanding the different mass case.} It was recently shown that at four loops a similar representation exists at higher orders in the dimensional regulator $\epsilon$, and this  naturally leads one to consider iterated integrals involving the canonical coordinate on the moduli space of a one-parameter family of Calabi-Yau three-folds~\cite{Pogel:2022ken}. These solutions and special functions, however, are not nearly as well studied as in the elliptic case, and a general understanding of the class of special functions that arise from Feynman integrals associated with Calabi-Yau geometries is currently still lacking, mostly due to a lack of explicit results for Feynman integrals.

The main goal of this paper is to present another example of an infinite family of multi-loop Feynman integrals associated to Calabi-Yau varieties, and to write down fully analytic results for them. More precisely, we consider the so-called $l$-loop \emph{ice cone graphs} in $d=2$ dimensions, obtained by inserting a $(l-1)$-loop banana integral into a one-loop three-point function with one massive external leg. All propagators are considered massive, and we assume that all propagator masses are equal. Remarkably, we find that the maximal cuts of the $l$-loop ice cone graph in $d=2$ dimensions organise into two copies of the same Calabi-Yau periods associated with the $(l-1)$-loop banana integrals. By analysing the system of differential equations for the ice cone family at low loop orders, we obtain a conjectural form of the differential equations at any number of loops. We find that this system is solved by the same class of special functions as in the case of the banana integrals, namely integrals involving the periods of the Calabi-Yau variety. This shows that the representation for the banana integrals obtained in ref.~\cite{Bonisch:2021yfw} defines an interesting class of iterated integrals that may describe a larger variety of multi-loop Feynman integrals. 

In the second part of this paper, we analyse how to write the banana and ice cone integrals as iterated integrals in the canonical coordinate on the moduli space. Using as a starting point the known mathematical literature on Calabi-Yau operators~\cite{bogner2013algebraic,Bognerthesis,MR3822913}, we find that at any number of loops the resulting iterated integrals are very similar to those obtained at four loops in ref.~\cite{Pogel:2022ken}, with the caveat that one needs to take into account more $Y$-invariants starting from six loops. Along the way, we show how it is possible to express the logarithmically-divergent periods themselves as iterated integrals of the same type, and how this representation allows one to interpret the quadratic relations among periods from Griffiths transversality as simple shuffle relations among iterated integrals.

This paper is organised as follows: In section~\ref{sec:icecones} we define the ice cone family. In section~\ref{sec:GM} we study the generalised leading singularities of the ice cone integrals, and we give a conjectural basis of master integrals at any number of loops. Our basis has the property that we can easily write down the system of differential equations satisfied by the master integrals for arbitrary loops. In section~\ref{subsec:mastersiterated} we apply the technique of ref.~\cite{Bonisch:2021yfw} to obtain solutions of the differential equations in terms of iterated integrals involving the periods of the Calabi-Yau variety. By studying the analytic structure of the solutions, we can fix the initial condition, and we conjecture a generating function for the integration constants. Finally, in section~\ref{sec:structure series} we show how one can change variables to the canonical coordinate on the moduli space, and we present fully analytic results in this form for both the banana and ice cone integrals. In section~\ref{sec:conclusion} we draw our conclusions. We include appendices where we collect known results for sub-topologies of the ice cone family, and where we present some technical proofs omitted in the main text.

%% file: icecone.tex
\section{The family of ice cone Feynman integrals}
\label{sec:icecones}

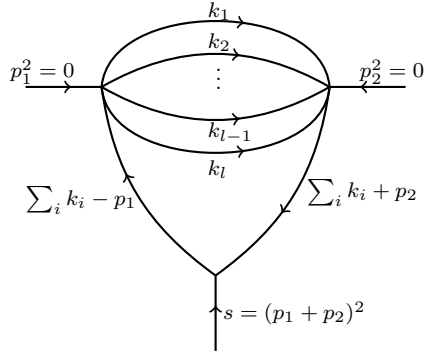
\begin{figure}[!h]
\centering
\begin{tikzpicture}
\coordinate (one) at (-2.5,0);
\coordinate (two) at (2.5,0);
\coordinate (One) at (-1.5,0);
\coordinate (Two) at (1.5,0);
\coordinate (Below) at (0,-2.5);
\coordinate (below) at (0,-3.5);
\begin{scope}[very thick,decoration={
    markings,
    mark=at position 0.6 with {\arrow{>}}}
    ] 
\draw [-, thick,postaction={decorate}] (one) to [bend right=0]  (One);
\draw [-, thick,postaction={decorate}] (two) to [bend right=0]  (Two);
\draw [-, thick,postaction={decorate}] (One) to [bend right=85]  (Two);
\draw [-, thick,postaction={decorate}] (One) to [bend right=30]  (Two);
\draw [-, thick,postaction={decorate}] (One) to [bend left=30]  (Two);
\draw [-, thick,postaction={decorate}] (One) to  [bend left=85] (Two);
\draw [-, thick,postaction={decorate}] (Below) to  [bend left=23] (One);
\draw [-, thick,postaction={decorate}] (Two) to  [bend right=-23] (Below);
\draw [-, thick,postaction={decorate}] (below) to  [bend left=0] (Below);
\end{scope}
\node (d1) at (0,1) [font=\scriptsize, text width=.2 cm]{$k_1$};
\node (d2) at (0,.6) [font=\scriptsize, text width=.2 cm]{$k_2$};
\node (p1) at (-2.2,.2) [font=\scriptsize, text width=1 cm]{$p_1^2=0$};
\node (p2) at (2.4,.2) [font=\scriptsize, text width=1 cm]{$p_2^2=0$};
\node (p3) at (1.3,-3) [font=\scriptsize, text width=2.4 cm]{$s=(p_1+p_2)^2 $};
\node (vd) at (.1,.2) [font=\scriptsize, text width=.2 cm]{$\vdots $};
\node (dlm1) at (0,-0.61) [font=\scriptsize, text width=.2 cm]{$k_{l-1}$};
\node (dl) at (0,-1.09) [font=\scriptsize, text width=.2 cm]{$k_{l}$};
\node (dlp1) at (-1.6,-1.5) [font=\scriptsize, text width=1.8 cm]{$\sum_i k_i-p_1$};
\node (dlp2) at (2.1,-1.4) [font=\scriptsize, text width=1.8 cm]{$\sum_i k_i +p_2$};
\end{tikzpicture}
\caption{The $l$-loop equal-mass ice cone graph build up from a $(l-1)$-loop banana graph and a one-loop triangle.}
\label{fig:icecone}
\end{figure}

We consider the  family of $l$-loop Feynman graphs obtained by inserting an $(l-1)$-loop banana graph into a one-loop three-point function, see figure~\ref{fig:icecone}. We will refer to this family as the 
\emph{ice cone graphs}.
The external momenta are such that $q=p_1+p_2$ with $p_1^2 = p_2^2 = 0$, and the graph only depends on
 $q^2=(p_1+p_2)^2 = s$ and on the propagator masses, which are all chosen to be equal to $m$.

At $l$ loops the graph contains $l+2$ propagators, which we parametrise as follows
\begin{equation}
\begin{aligned}
	D_a	&=	k_a^2-m^2\, ,	\quad 1\le a\le l\, ,	\\
	D_{l+1}	&=	\big(\sum_i k_i-p_1\big)^2-m^2	\,, \quad&\quad	
    D_{l+2}	&=	\big(\sum_i k_i+p_2\big)^2-m^2\, ,	
\label{defprops}
\end{aligned}
\end{equation}
where the first $l$ propagators define the $(l-1)$-loop banana graph, while the last two
come from the one-loop triangle.
 We can form $l(l+5)/2$ independent scalar products between the $l$ loop momenta and the two independent external ones, 
and so we need to  supplement the propagators with 
$n = (l-1)(l+4)/2$ extra propagators, or equivalently irreducible scalar products. While the choice made here is in principle arbitrary,
a good choice of numerators can help to simplify the computation.
Here it is convenient to choose the  numerators $N_r$ ($r=1,...,n$) as follows:
\begin{align}
N_1	&=	\big(\sum_i k_i\big)^2\,,  \nonumber \\
N_r &= \left\{ \begin{array}{ccc}  	
         k_i\cdot k_j       & \hspace{0.7cm} \text{for } i=1,\hdots,j-1, &  \ j=3,\hdots, l \,,	\\
	k_i\cdot p_1	&  \text{for } i=2,\hdots, l\,, &	\\
	k_i\cdot p_2	& \text{for } i=2,\hdots, l\,, &   \end{array} \right. \qquad \mbox{for} \quad  2\leq r \leq n\,.
\label{eq:isps}	
\end{align}
A generic integral in the ice cone family then takes the form
\begin{equation}
\begin{aligned}
	I_{\underline \nu}(s;d)	=	
	\int\left( \prod_{j=1}^l \frac{\mathrm d^dk_i}{i\pi^{d/2}}	\right) 
	\frac{N_1^{-\nu_{l+3}}N_2^{-\nu_{l+4}} \cdots N_n^{-\nu_{l(l+5)/2}}}{D_1^{\nu_1} \cdots D_{l+2}^{\nu_{l+2}}}	\, ,
\end{aligned}
\end{equation}
where $\underline{\nu} = (\nu_1,...,\nu_{l+2} ; \nu_{l+3},..., \nu_{l(l+5)/2} ) $.
In what follows, we will be particularly interested in integrals where only the first two scalar products appear with
powers greater than zero. We will therefore use a shorthand notation where we drop trailing zero entries: 
\beq
\underline{\nu} = (\nu_1,...,\nu_{l+2} ; \nu_{l+3},\nu_{l+4}, 0,..., 0 ) = (\nu_1,...,\nu_{l+2} ; \nu_{l+3},\nu_{l+4})\,,
\eeq
and we just write
\begin{equation}
\begin{aligned}
	I_{\underline \nu}(s;d)	=	
	\int\left( \prod_{j=1}^l \frac{\mathrm d^dk_i}{i\pi^{d/2}}	\right) 
	\frac{N_1^{-\nu_{l+3}}N_2^{-\nu_{l+4}}}{D_1^{\nu_1} \cdots D_{l+2}^{\nu_{l+2}}}	\, .
\label{defintfam}
\end{aligned}
\end{equation}
As it will become clear below, at any number of loops, the first two numerators are sufficient to define a basis of master integrals for which the symmetries of the problem become manifest. 
Notice that we are of course always free to set $m=1$, such that all integrals are only functions of the single variable $s$. We will do this later during the explicit computation of the master integrals. For now, we keep exact dependence on $m$, as it makes the structure of some formulas clearer.

\section{Differential equations for the ice cone family}
\label{sec:GM}
In this section we study the system of differential equations satisfied by the 
master integrals of the ice cone family at an arbitrary number of loops. Following the
mathematical literature on the subject, we will refer to this system as the
\emph{Gauss-Manin} (GM) system of differential equations. 
By studying what we call the \emph{generalised leading singularities} (GLS)
associated to the ice cone graph at $l$-loops in $d=2$ space-time dimensions, 
we will show that the master integrals can be organised into two {disjoint sets}, corresponding to two
equivalent replicas of the $(l-1)$-loop banana graph, plus
one extra master integral, that can always be chosen to be equal to 
zero in strictly $d=2$ dimensions.

\subsection{A basis of generalised leading singularities for the ice cone graph}
\label{sec:gls}

We consider the master integrals for the $l$-loop ice cone family in $d=2 -2 \epsilon$
space-time dimensions, knowing that the corresponding results in $d=4 -2 \epsilon$ can be obtained
from our calculation by a dimensional shift~\cite{Tarasov:1996br,Lee:2009dh}.
Let us consider the so-called \emph{corner integral} in the top sector of the ice cone family, which in our notation reads:
\begin{align}
I_{1,...,1;0,0}(s;d)	=	
	\int\left( \prod_{j=1}^l \frac{\mathrm d^dk_i}{i\pi^{d/2}}	\right) 
	\frac{1}{D_1\cdots D_{l+2}}	\, .
\end{align}
By studying the residues and the branch cuts of the integrand above, we can infer
 information about the number of master integrals appearing in this problem and the type of functions
required for their calculation~\cite{Lee:2013hzt,Georgoudis:2016wff,Primo:2016ebd,Frellesvig:2017aai,Bosma:2017ens,Primo:2017ipr,Bitoun:2018afx,Mastrolia:2018uzb,Caron-Huot:2021xqj}.
To this end, it is convenient to write the integral in a way that separates the integration over the two triangle propagators
from the ones over the internal $(l-1)$-loop banana sub-graph. By changing integration
variables to $\sum_{i=1}^{l} k_i = k$, we write the ice cone integral as
\begin{equation}
I_{1,...,1;0,0}(s;d) = \int \frac{\mathrm d^d k}{i\pi^{d/2}} \frac{{\rm Ban}^{(l-1)}(k^2)}{((k-p_1)^2-m^2)((k+p_2)^2-m^2)} \, ,
\end{equation}
where ${\rm Ban}^{(l-1)}(k^2)$ is the $(l-1)$-loop banana sub-graph evaluated at the external momentum squared $k^2= N_1$.
To study cuts and residues of the integral, it is convenient to parametrise the integrand 
using Baikov's change of variables~\cite{Baikov:1996rk,Baikov:1996iu,Harley:2017qut,Frellesvig:2017aai}, 
which promotes the propagators to integration variables.
We define:
\beq
 z_1 = ((k-p_1)^2-m^2)\,,\qquad z_2 = (k+p_2)^2-m^2\,, \qquad z_3 =k^2\,.
 \eeq
The integral then becomes, up to an irrelevant overall function of $d$,
\begin{equation}
I_{1,...,1,0,0}(s;d) \propto \int_\mathcal{C} \rd z_1 \rd z_2 \rd z_3 \frac{ P(s,m^2,z_1,z_2,z_3)^{(d-4)/2}}{z_1 z_2} {\rm Ban}^{(l-1)}(z_3) \, ,
\label{eq:baikov}
\end{equation}
where the Baikov polynomial reads
\begin{align}
P(s,m^2,z_1,z_2,z_3) = m^4 + m^2(z_1 + z_2 -2 z_3) + z_1(z_2 -z_3) + z_3(s-z_2+z_3) \,,
\end{align}
and the details of the integration contour $\mathcal{C}$ will be irrelevant in what follows.

We start by fixing $d=2$ in eq.~\eqref{eq:baikov}.\footnote{It is possible that using methods based on twisted cohomology theory, this could be generalised to $d=2-2 \epsilon$ dimensions, see for example ref.~\cite{Giroux:2022wav}. 
For our scopes, an analysis in $d=2$ is sufficient.}
Moreover, we consider the integrand associated to the so-called maximal cut, which is obtained by cutting
the two triangle propagators and the banana graph, i.e., by putting them on shell. This means that our analysis will not be sensitive to the sub-graphs obtained by pinching any propagators of the ice cone graph. 
We deform the integration contour to encircle the two poles at $z_1=0$ and $z_2=0$, and we obtain
\begin{equation}
{\rm Cut} \left[ I_{1,...,1;0,0}(s;2) \right] 
\propto \oint \frac{\rd z_3}{  z_3^2+ (s-2m^2)z_3 + m^4} {\rm Cut} \left[ {\rm Ban}^{(l-1)}(z_3) \right] \, .
\label{eq:cutbaikov}
\end{equation}
The analytic structure of the integrand becomes manifest when reparametrising $s$ through the 
so-called Landau variable
\begin{equation}
  s = -m^2 \frac{(1-x)^2}{x}\,,
\label{eq:landau}
\end{equation}
such that the maximal cut integral becomes 
\begin{equation}
{\rm Cut} \left[ I_{1,...,1;0,0}(s;2) \right] 
\propto \oint \frac{\rd z_3}{  \left( z_3-m^2 x \right) \left( z_3 - \frac{m^2}{x} \right)} {\rm Cut} 
\left[ {\rm Ban}^{(l-1)}(z_3) \right]\, .
\label{eq:baikov2}
\end{equation}
The next step is to analyse this integrand to determine
its \emph{generalised leading singularities} (GLS), defined as the
integrals associated to the different integration contours (or cycles) that provide
independent results. This is a natural generalisation of the concept of leading singularities of Feynman integrals, defined as an algebraic function obtained by taking a maximal codimension residue of the integrand~\cite{Cachazo:2008vp}. Various extensions of this concept to include cases where it is not possible to localise the integrand by taking residues have appeared in the literature, cf., e.g., refs.~\cite{Primo:2017ipr,Primo:2016ebd,Bosma:2017ens,Bonisch:2021yfw,Bourjaily:2020hjv,Bourjaily:2021vyj,Bourjaily:2022tep}, and ours seamlessly fits into these definitions.
The GLS will furnish a basis of solutions for the homogeneous differential equations
satisfied by the ice cone graph~\cite{Primo:2016ebd,Primo:2017ipr}. Moreover, as we will see, their
analysis will provide a way to determine a good basis of master integrals,
whose Gauss-Manin differential equation assumes a particularly convenient block-diagonal form.

We start by noticing that the integrand in
eq.~\eqref{eq:baikov2} has two simple poles at $z_3 = m^2x$ and $z_3 = m^2/x$. The
two residues associated with these poles define the GLS. From eq.~\eqref{eq:baikov2},
we notice that they can be evocatively written in terms 
of the corresponding maximal cuts of the $(l-1)$-loop banana graph 
computed at special values of its argument
\begin{equation}
 {\rm Cut} 
\left[ {\rm Ban}^{(l-1)}(m^2 x) \right]\,,\qquad 
{\rm Cut} 
\left[ {\rm Ban}^{(l-1)}(m^2/x) \right]
\,.
\end{equation}
It is known that each maximal cut of the $(l-1)$-loop banana graph 
has in turn $(l-1)$ independent GLS, which can be expressed in terms of the independent 
periods of a $(l-2)$-dimensional Calabi-Yau
manifold~\cite{Klemm:2019dbm,Bonisch:2020qmm,Bonisch:2021yfw,Vanhove:2014wqa,Vanhove:2018mto,Bloch:2014qca,Bloch:2016izu}. We get in this way $2(l-1)$
independent GLS in $d=2$ space-time dimensions, organised into two copies of the periods of the same
$(l-2)$-dimensional Calabi-Yau manifold appearing in the computation of the banana graphs.
From this, and remembering that $z_3 = N_1$ in eq.~\eqref{eq:isps}, we can easily choose master integrals with numerators that `disentangle' these two copies of the banana graph in $d=2$.
In this way, we are lead to choose the master integrals as follows
\beq\label{eq:mastersdis}
\widetilde\cI_{l,k}^{\pm}(x;2) = -m^2 x^{\pm1}\, I_{\underbrace{\scriptstyle2,\ldots,2}_{k},1,\ldots,1;0,0}(s;2)  + I_{\underbrace{\scriptstyle2,\ldots,2}_{k},1,\ldots,1;-1,0}(s;2)\,,\qquad 1\le k<l\,.
\eeq

This analysis has been performed in $d=2$ dimensions. We will see below that, once we consider
Integration-By-Parts (IBP) relations~\cite{Chetyrkin:1981qh,Tkachov:1981wb} in $d$ dimensions, the maximal cuts of the ice cone graph
turn out to have one more master integral 
compared to the naive analysis above, i.e.,
there are in total $2l-1$ independent master integrals.
The fact that we do not see this extra master integral in our analysis suggests that it must be possible to choose it to be identically zero in $d=2$. In other words, only $2(l-1)$ masters integrals in the top sector can be linearly independent in $d=2$ dimensions.
To find this extra master integral, we use the construction proposed in ref.~\cite{Remiddi:2013joa}\footnote{For similar considerations in the context of massless integrals, see also ref.~\cite{Gluza:2010ws}.} and consider
the following integral in $d$ space-time dimensions
\begin{align}
    \widetilde\cI_{l,l}(x;d) = \int \frac{\rd^d k}{i \pi^{d/2}} \frac{G(k,p_1,p_2)}{((k-p_1)^2-m^2)((k+p_2)^2-m^2)} \,{\rm Ban}^{(l-1)}(k^2)\,,
    \label{eq:extraI}
\end{align}
where $G(k,p_1,p_2)$ is the Gram determinant of the three momenta, also referred to as Schouten polynomial in ref.~\cite{Remiddi:2013joa},
\begin{equation}
    G(k,p_1,p_2) 
    = \frac{s^2}{4} k^2 - s (k \cdot p_1)(k \cdot p_2) 
\,.
\label{eq:gram}
\end{equation}
There is nothing special about eq.~\eqref{eq:extraI} in  $d$ dimensions, and nothing forbids us from choosing it as an element of our
basis of master integrals for the ice cone family.
On the other hand, it is easy to see that in $d=2$
dimensions this integral does not have any IR divergences and, despite the numerator, it remains UV convergent. Since the Gram determinant
of three momenta in two dimensions is zero, the integral must itself be identically zero in $d=2$:
\beq
\label{eq:I_l,l=0}
\cI_{l,l}(x,d) = \ord(d-2)\,.
\eeq

Let us conclude this section by illustrating our choice of basis of master integrals in $d=2-2\epsilon$ at low loop order. We start with the case of the two-loop ice cone graph.
Using the fact that the maximal cut of the one-loop bubble of momentum $p^2$ 
in $d=2$
can be written as
\beq
{\rm Cut} 
\left[ {\rm Ban}^{(1)}(p^2;2) \right] \propto \frac{1}{\sqrt{p^2(p^2-4 m^2)}} \, ,
\eeq
we see that eq.~\eqref{eq:baikov2} simplifies to
\begin{equation}
{\rm Cut} \left[ I_{1,1,1,1;0,0}(s;2) \right] 
\propto \oint 
\frac{\rd z_3}{  \left( z_3-m^2 x \right) \left( z_3 - \frac{m^2}{x} \right)}\frac{1}{\sqrt{z_3(z_3-4 m^2)}} \,,
\label{eq:twoloops}
\end{equation}
 and the two GLS read
 \begin{align}
     -\frac{x}{m^4 (1-x^2)\sqrt{x(x-4)}}\,, \qquad \frac{x^2}{m^4(1-x^2)\sqrt{1-4x}}
\,,
 \end{align}
 such that  at two loops we expect two master integrals in $d=2$ dimensions in the top sector. 
 A reduction to master integrals using, 
 for example, \texttt{Reduze2}~\cite{Studerus:2009ye,vonManteuffel:2012np} instead reveals 
 three master integrals in $d$ dimensions in the top sector, 
 plus three simpler master integrals for the sub-topologies. First of all, the integrals in the subtopologies
 are trivial tadpoles and bubbles and, for convenience, we choose finite candidates for them in two dimensions. 
 We therefore define  the remaining master integrals as follows\footnote{Here we set $m=1$ for simplicity, 
 and since basis changes like \eqref{eq:rotl3} are trivial for ${\widetilde {\cal I}}_{l,1}^\pm$ for all 
 $l$ we do not to distinguish between ${\cal I}_{l,1}^\pm$ and ${\widetilde {\cal I}}^\pm_{l,1}$ below.} :
 \begin{equation}
 \begin{aligned}
 	I_{0,0}(x;d)	&=	I_{2,2,0,0;0,0}(s;d) \, ,	\\	
    I_{0,1}(x;d)	&=	I_{1,1,1,0;0,0}(s;d)  \,,	\\
    I_0(x;d)	    &=	I_{2,0,1,1;0,0}(s;d)  \,, \\
    \cI_{2,1}^+(x;d)	    &=	-x I_{1,1,1,1;0,0}(s;d) +  I_{1,1,1,1;-1,0}(s;d) \,,	\\
    \cI_{2,1}^-(x;d) 	    &=	-\frac{1}{x} I_{1,1,1,1;-1,0}(s;d) +  I_{1,1,1,1;-1,0}(s;d)\, ,	\\
	\cI_{2,2}(x;d)	    &=	\frac{1}{4} I_{1,1,1,0;0,0}(s;d) + \frac{1}{4} I_{1,1,1,1;0,0}(s;d) - \frac{1-x+x^2}{4x} I_{1,1,1,1;-1,0}(s;d) \\
                    &\quad + I_{1,1,1,1;-1,-1}(s;d) \, ,
\label{eq:mis2loop}
\end{aligned}
\end{equation}
 where $I_{0,0}(x;d)$, $I_{0,1}(x;d)$ and $I_0(x;d)$ are sub-topologies, while the linear combination that defines $ \cI_{2,2}(x;d)	$, is obtained
 by starting from eq.~\eqref{eq:extraI}, and reducing it to the standard master integrals for the problem. 
By a similar reasoning, we find the following basis of master integrals at three loops:
\begin{align}
\nonumber	I_{0,0}(x;d)	&=	I_{2,2,2,0,0;0,0}(s;d) \, ,	\\	
 \nonumber   I_{0,1}(x;d)	&=	I_{1,1,1,1,0;0,0}(s;d) \, ,	\\
\label{eq:defmastersl3}	I_0(x;d)	    &=	I_{2,2,0,1,1;0,0}(s;d) \, , \\
\nonumber	\widetilde\cI_{3,1}^+(x;d)	    &=	-x I_{1,1,1,1,1;0,0}(s;d) +  I_{1,1,1,1,1;-1,0}(s;d)\, ,	\\
\nonumber \widetilde\cI_{3,2}^+(x;d)	  &  =	-x I_{2,1,1,1,1;0,0}(s;d) +  I_{2,1,1,1,1;-1,0}(s;d)\, ,	\\
\nonumber    \widetilde\cI_{3,1}^-(x;d)	     &=  -\frac1x I_{1,1,1,1,1;0,0}(s;d) +  I_{1,1,1,1,1;-1,0}(s;d) \,, \\ 
\nonumber    \widetilde\cI_{3,2}^-(x;d)	     &=  -\frac1x I_{2,1,1,1,1;0,0}(s;d) +  I_{2,1,1,1,1;-1,0}(s;d)\, , \\
\nonumber	\cI_{3,3}(x;d)		    &=	\frac16 I_{1,1,1,1,0;0,0}(s;d) + \frac16 I_{1,1,1,1,1;0,0}(s;d) - \frac{1-x+x^2}{6x} I_{1,1,1,1,1;-1,0}(s;d)\\
\nonumber&\quad + I_{1,1,1,1,1;-1,-1}(s;d) \, .
\end{align}

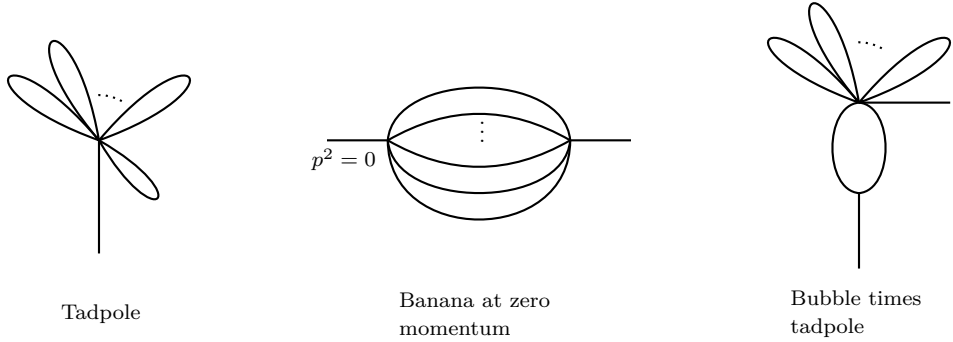
\begin{figure}[!t]
\centering
\begin{tikzpicture}
\coordinate (D1One) at (-5.0,-1.5);
\coordinate (D1Two) at (-5.0,0);
\coordinate (D2one) at (-2.0,0);
\coordinate (D2two) at (2.0,0);
\coordinate (D2One) at (-1.2,0);
\coordinate (D2Two) at (1.2,0);
\coordinate (D3one) at (5,-1.7);
\coordinate (D3One) at (5,-.7);
\coordinate (D3two) at (6.2,.5);
\coordinate (D3Two) at (5,.5);
\centerarc[dotted,thick](-5.0,0)(60:95:.6) ;
\centerarc[dotted,thick](5.0,.7)(60:95:.6) ;
\begin{scope}[thick,decoration={
    markings,
    mark=at position 0.6 with {}}
    ] 
\draw [-, thick,postaction={decorate}] (D1One) to [bend right=0]  (D1Two);  
 \foreach \angle in {20}
\draw   [-, thick,postaction={decorate}] (D1Two) .. controls +(\angle+30:2cm) and +(\angle:2cm) .. (D1Two); 
\foreach \angle in {100, 130}
 \draw   [-, thick,postaction={decorate}] (D1Two) .. controls +(\angle+30:2cm) and +(\angle:2cm) .. (D1Two); 
  \draw   [-, thick,postaction={decorate}] (D1Two) .. controls +(-30:1.5cm) and +(-60:1.5cm) .. (D1Two); 
\draw [-, thick,postaction={decorate}] (D2one) to [bend right=0]  (D2One);
\draw [-, thick,postaction={decorate}] (D2Two) to [bend right=0]  (D2two);
\draw [-, thick,postaction={decorate}] (D2One) to [bend right=85]  (D2Two);
\draw [-, thick,postaction={decorate}] (D2Two)   .. controls +(0,-1.4) and +(0,-1.4) ..  (D2One);
\draw [-, thick,postaction={decorate}] (D2One) to [bend right=30]  (D2Two);
\draw [-, thick,postaction={decorate}] (D2One) to [bend left=30]  (D2Two);
\draw [-, thick,postaction={decorate}] (D2One) to  [bend left=85] (D2Two);
\draw [-, thick,postaction={decorate}] (D3one) to [bend right=0]  (D3One);
\draw [-, thick,postaction={decorate}] (D3Two) to [bend right=0]  (D3two);
\draw [-, thick,postaction={decorate}] (D3One) to [bend right=-85]  (D3Two);
\draw [-, thick,postaction={decorate}] (D3Two) to  [bend left=85] (D3One);
\foreach \angle in {20}
\draw   [-, thick,postaction={decorate}] (D3Two) .. controls +(\angle+30:2cm) and +(\angle:2cm) .. (D3Two); 
\foreach \angle in {100, 130}
 \draw   [-, thick,postaction={decorate}] (D3Two) .. controls +(\angle+30:2cm) and +(\angle:2cm) .. (D3Two); 
 \end{scope};
\node (vd) at (.1,.2) [font=\scriptsize, text width=.2 cm]{$\vdots $};
\node (d0) at (-1.7,-0.25) [font=\scriptsize, text width=1 cm]{$p^2=0$};
\node (d1) at (-5.4,-2.3) [font=\scriptsize, text width=.2 cm]{Tadpole};
\node (d2) at (-.05,-2.3) [font=\scriptsize, text width=2 cm]{Banana at zero momentum};
\node (d3) at (5.1,-2.3) [font=\scriptsize, text width=2 cm]{Bubble times tadpole};
\end{tikzpicture}
\caption{Sub-topologies of the ice cone graph with corresponding master integrals $I_{0,0}$ (left), $I_{0,1,},\hdots, I_{0,\lfloor l/2 \rfloor}$ (middle) and $I_0$ (right).}
\label{fig:degenerations}
\end{figure}

From these two- and three-loop examples we see a pattern emerging, and we claim that this pattern continues to higher loops.
We conjecture that for the general $l$-loop ice cone family the following integrals
form a suitable basis of master integrals (cf. eq.~\eqref{eq:mastersdis}):
\allowdisplaybreaks
\begin{align}
	I_{0,0}	&=	I_{2,\hdots,2,0,0;0,0}  , \nonumber	\\ 	
 I_{0,k} &=  I_{\underbrace{\scriptstyle 2,\hdots,2}_{2(k -1)}\hspace{-0.3ex},1,\hdots,1,0;0,0}\,,\qquad 1\le k\le \lfloor l/2 \rfloor\,  , \nonumber   \\
    I_0	    &=	I_{2,\hdots,2,0,1,1;0,0}  , \nonumber \\
\label{eq:defmastersll}
\widetilde\cI_{l,k}^{+}& = -m^2 x\, I_{\underbrace{\scriptstyle2,\ldots,2}_{k},1,\ldots,1;0,0}  + I_{\underbrace{\scriptstyle2,\ldots,2}_{k},1,\ldots,1;-1,0}\,,\qquad 1\le k<l\,,\\
\nonumber\widetilde\cI_{l,k}^{-} &= -\frac{m^2}{x}\, I_{\underbrace{\scriptstyle2,\ldots,2}_{k},1,\ldots,1;0,0}  + I_{\underbrace{\scriptstyle2,\ldots,2}_{k},1,\ldots,1;-1,0}\,,\qquad 1\le k<l\,,\\
\nonumber	\cI_{l,l}	    &=	\frac1{2l} I_{1,\hdots,1,0;0,0} + \frac1{2l} I_{1,\hdots,1;0,0} - \frac{1-x+x^2}{2lx} I_{1,\hdots,1;-1,0} + I_{1,\hdots,1;-1,-1} \, .
\end{align}
Besides the master integrals $\cI_{l,k}^{\pm}$ and $\cI_{l,l}$ from the top sector, there are $\lfloor l/2\rfloor+2$ master integrals from the sub-sectors (see figure~\ref{fig:degenerations}). We have explicitly checked that this is true up to five loops, using both \texttt{Reduze2}~\cite{Studerus:2009ye,vonManteuffel:2012np} and \texttt{Kira2}~\cite{Maierhofer:2017gsa,Klappert:2020nbg}. For the six-loop family we have also verified that there are three masters $I_{0,1},I_{0,2},I_{0,3}$  
for the banana sub-topology at zero momentum (middle graph in figure \ref{fig:degenerations}).\footnote{We could not verify this patter to higher loops due to the intrinsic limitation in both reduction programs  to a maximum of $32$ propagators in an integral family.} 
We stress that, while the analysis of the cuts was performed in $d=2$ dimensions, the basis in eq.~\eqref{eq:defmastersll} is valid in $d$ dimensions.

We conclude this section by mentioning that the corner integral $I_{1,\hdots,1;0,0}(s;d)$ is to all loop orders
given  by the simple linear combination
\begin{equation}
\begin{aligned}
	I_{1,\hdots,1;0,0}(s;d) =   \frac{x}{1-x^2} \left[\widetilde{\cI}_{l,1}^+(x;d)- \widetilde{\cI}_{l,1}^-(x;d)\right]   \, .
\label{eq:origice}
\end{aligned}
\end{equation}
At this point we make an observation that will be important later on. The definition of the Landau variable in eq.~\eqref{eq:landau} is invariant under $x\to 1/x$. The Feynman integral $I_{1,\hdots,1;0,0}(s;d)$ is then also invariant: it is \emph{parity-even}. Since the rational prefactor in eq.~\eqref{eq:origice} is parity-odd, we see that we the following relation must be satisfied
\beq\label{eq:parity}
\widetilde{\cI}_{l,1}^-(x;d) = \widetilde{\cI}_{l,1}^+(1/x;d)\,.
\eeq

\subsection{Gauss-Manin differential system for the ice cone family}
\label{sec:lloop}
It is well known that the vector of master integrals satisfies a homogenous first-order differential equation~\cite{Kotikov:1990kg,Kotikov:1991hm,Kotikov:1991pm,Gehrmann2000,Henn:2013pwa}, which we will refer to as the \emph{Gauss-Manin} (GM) system in the following. In this section we will focus on its properties in strictly $d=2$ dimensions. Our aim is to argue that, just like for the basis of master integrals, we can {conjecturally} write down the GM system for arbitrary loop order.

Before we discuss the general $l$-loop case, let us focus on $l=2,3$. Our basis of master integrals was presented at the end of the previous section. For $l=2$, we consider the vector of master integrals $\underline I^{(2)}=(I_{0,0},I_{0,1},I_0, \cI_{2,1}^+	, \cI_{2,1}^-, \cI_{2,2})$. We also shorten our notation by leaving the dependence of our master integrals
 on $x$ and $d$ implicit, where $x$ is the Landau variable 
 introduced in eq.~\eqref{eq:landau}. With this, one finds by direct calculation that the GM system takes the following form:
\begin{equation}
\begin{aligned}
	\frac{\mathrm d}{\mathrm dx}\underline I^{(2)} =  	\left(
		\begin{array}{cccccc}
            0 & 0 & 0 & 0 & 0 & 0 \\
            0 & 0 & 0 & 0 & 0 & 0 \\
            -\frac{2}{(1-x) (1+x)} & 0 & \frac{1+x^2}{x(1-x)(1+x)} & 0 & 0 & 0 \\
            0 & 0 & \frac{(1-x) (1+x)}{x (1-4 x)} & \frac{1-2 x}{x (1-4 x)} & 0 & 0 \\
            0 & 0 & -\frac{(1-x) (1+x)}{x^2 (4-x)} & 0 & -\frac{2-x}{x (4-x)} & 0 \\
            0 & 0 & 0 & 0 & 0 & 0 \\
        \end{array}
	\right)	\underline I^{(2)} + \mathcal{O}(d-2)\, .
\label{eq:gm2}
\end{aligned}
\end{equation}
 Equation~\eqref{eq:gm2} has the desired block-triangular form
 in $d=2$, with only three non-trivial blocks on the diagonal. Moreover, the master integral $\cI_{2,2}$
 is completely decoupled from the system of differential equations for $d=2$, in agreement with eq.~\eqref{eq:I_l,l=0}. Focusing on the master integrals from the top sector, we can write eq.~\eqref{eq:gm2} in the equivalent form of a system of inhomogeneous equations:
\beq\bsp
\frac{\mathrm d}{\mathrm dx} \cI^{+}_{2,1}&\,= \frac{1-2 x}{x (1-4 x)}\, \cI^{+}_{2,1} + \frac{(1-x) (1+x)}{x (1-4 x)}\, I_0 + \ord(d-2)\,,\\
\frac{\mathrm d}{\mathrm dx} \cI^{-}_{2,1}&\,= -\frac{2-x}{x (4-x)}\, \cI^{-}_{2,1} - \frac{(1-x) (1+x)}{x^2 (4-x)}\, I_0+\ord(d-2)\,,\\
\frac{\mathrm d}{\mathrm dx} \cI_{2,2}&\,= \ord(d-2)\,.
\esp\eeq
We see that the equations in the top sector completely decouple, and we can easily solve for $\cI^{+}_{2,1}$, $\cI^{-}_{2,1}$ and $\cI_{2,2}$.

 For $l=3$, we derive the GM system for the master integrals in eq.~\eqref{eq:defmastersl3}. The GM system has a simple form in two dimensions, where the two two-loop banana blocks are visible on the diagonal. 
We put the equations in a more convenient form by performing an additional rotation only on these two blocks, such that the 
resulting equations have as solutions exactly the maximal cuts of the two-loop banana integral and its derivative in $x$. This means that we choose as basis elements exactly linear combinations of integrals that correspond to the various derivatives with respect to $x$
of the banana graph inserted into the ice cone graph. While at this point there is no physical reason to prefer this basis over any other, our choice makes it 
easier in practice to compute all master integrals from the corner integral with unit propagator powers. The explicit form of this rotation is given by
\begin{equation}
\begin{aligned}
	\begin{pmatrix}
	   \cI_{3,1}^+ \\ \cI_{3,2}^+ \\ \cI_{3,1}^- \\ \cI_{3,2}^-
	\end{pmatrix}  =   %
    \left(
    \begin{array}{cccc}
        1 & 0 & 0 & 0 \\
        \frac{1}{x} & \frac{3}{x} & 0 & 0 \\
        0 & 0 & 1 & 0 \\
        0 & 0 & -\frac{1}{x} & -\frac{3}{x} \\
    \end{array}
    \right) %
    \begin{pmatrix}
        \widetilde\cI_{3,1}^+ \\ \widetilde\cI_{3,2}^+ \\ \widetilde\cI_{3,1}^- \\ \widetilde\cI_{3,2}^-
	\end{pmatrix}  \, .
\label{eq:rotl3}
\end{aligned}
\end{equation}
Our final basis of master integrals is then $\underline I^{(3)}=(I_{0,0},I_{0,1},I_0,\cI_{3,1}^+,\cI_{3,2}^+,\cI_{3,1}^-,\cI_{3,2}^-,\cI_{3,3})$. The GM system then takes the form:
\begin{align}
\label{eq:gm3}	\frac{\mathrm d}{\mathrm dx}\underline I^{(3)} &= \\	
\nonumber		&\hspace{-8ex}\left(
		\begin{array}{cccccccc}
			 0 & 0 & 0 & 0 & 0 & 0 & 0 & 0 \\
			 0 & 0 & 0 & 0 & 0 & 0 & 0 & 0 \\
 			-\frac{2}{(1-x) (1+x)} & 0 & \frac{1+x^2}{x(1-x)  (1+x)} & 0 & 0 & 0 & 0 & 0 \\
			 0 & 0 & 0 & 0 & 1 & 0 & 0 & 0 \\
			 0 & 0 & \frac{3 (1+x)}{x^2 (1-9 x)} & -\frac{1-3 x}{x^2 (1-x) (1-9 x)} & \frac{(1-3 x) (1+3 x)}{x (1-x) (1-9 x)} & 0 & 0 & 0	 \\
 			0 & 0 & 0 & 0 & 0 & 0 & 1 & 0 \\
 			0 & 0 & -\frac{3 (1+x)}{x^2 (9-x)} & 0 & 0 & \frac{3-x}{x (1-x) (9-x)} & -\frac{9-20x+3 x^2}{x (1-x) (9-x)} & 0 \\
 			0 & 0 & 0 & 0 & 0 & 0 & 0 & 0 \\
	   \end{array}
	   \right)	\underline I^{(3)}\\
 \nonumber       &\hspace{11.5cm}+ \mathcal O(d-2)\, .
\end{align}
We can again focus on the top sector and look at the corresponding inhomogeneous equations:
\beq\bsp
\frac{\rd}{\rd x}\,\underline\cI_{3}^+&\,={\bf GM}^{(2)}_{\textrm{ban}}(x)\,\underline\cI_{3}^+ + \underline{{N}}_3^{+}\,I_0+\ord(d-2)\,,\\
\frac{\rd}{\rd x}\,\underline\cI_{3}^-&\,={\bf GM}^{(2)}_{\textrm{ban}}(1/x)\,\underline\cI_{3}^- + \underline{{N}}_3^{-}\,I_0+\ord(d-2)\,,\\
\frac{\rd}{\rd x}\,\,\cI_{3}&\,=0\,,
\esp\eeq
where we defined $\underline\cI_{3}^{\pm} = (\cI_{3,1}^\pm,\cI_{3,2}^\pm)^T$ and
\beq\bsp 
\underline{{N}}_3^{+} = \left(0,\frac{3 (1+x)}{x^2 (1-9 x)}\right)^T \,, \quad \underline{{N}}_3^{-} = \left(0,-\frac{3 (1+x)}{x^2 (9- x)}\right)^T\,.
\esp\eeq
At this point we make an important observation.
The entries of the matrix ${\bf GM}^{(2)}_{\textrm{ban}}$ can easily be read off from eq.~\eqref{eq:gm3}. It is precisely the matrix that describes the GM system satisfied by the maximal cuts of the two-loop banana integral! We can write with our basis choice:
\beq\label{eq:3-loop_derivative}
\cI_{3,2}^{\pm} = \frac{\rd}{\rd x}\cI_{3,1}^{\pm} \,.
\eeq
Moreover, we see that, just like at two loops, the GM system for the top sector splits into three independent blocks that can be solved independently, and  the inhomogeneities only involve the integral $I_0$, but not $I_{0,k}$. 

We conjecture that such a simple structure for the GM system persists through all loop orders. One again has to include an additional rotation such that the two 
diagonal blocks associated to $\underline\cI_{l}^{\pm}$ are solved by the maximal cuts of the $(l-1)$-loop banana integral and its derivatives.
 A general form of this transformation 
is hard to write down, because there is no known closed form for the GM system for banana integrals at $l$ loops. Nevertheless, as explained in refs.~\cite{Vanhove:2014wqa,Bonisch:2020qmm}, 
a simple algorithm to produce them exists. Note that at all loop orders $l$ 
the integrals $\widetilde\cI_{l,1}^{\pm}$ are not modified by the additional rotation displayed for $l=3$ in~\eqref{eq:rotl3}. We have again checked that our claim holds to at least five loops using {\tt Reduze2} and {\tt Kira2}.

In the remainder of this section we describe our conjectural form for the GM system to all loop orders. 
In the rotated basis $\underline I^{(l)}=(I_{0,0},I_{0,1},\hdots,I_{0,\lfloor l/2 \rfloor},I_0,\underline\cI^+_l,\underline\cI^-_l,\cI_{l,l})^T$, the Gauss-Manin system has just three non-trivial blocks. 
The master integrals $\cI_{l,l}$ and $I_{0,k}$, $0\le k\le\lfloor l/2 \rfloor$, satisfy trivial
differential equations, i.e., they are just constants in $d=2$ dimensions. We know from eq.~\eqref{eq:I_l,l=0} that $\cI_{l,l}$ vanishes in $d=2$ dimensions. 
Results in closed analytic form are easy to obtain for 
the master integrals $I_{0,0}$ and $I_{0}$ at an arbitrary number 
of loops. In particular, the integral $I_{0,0}$ is a product of (convergent) one-loop tadpole integrals, and it evaluates to unity in $d=2$ dimensions (see appendix~\ref{app:derivationmaster}). 
The remaining integrals $I_{0,k}$, $1\le k\le \lfloor l/2 \rfloor$, are $l$-loop banana integrals evaluated at zero external momentum. Their  numerical values cannot be expressed in terms of known transcendental numbers for loop orders $l>3$. 
Nevertheless, there exist numerically fast convergent integral representations for them in terms of Bessel functions~\cite{Bonisch:2020qmm,Bonisch:2021yfw,bailey2008elliptic,Broadhurst:2016myo,Broadhurst2013}. 
As we will see below, these constants are not required to solve the differential equation for the corner integral 
in $d=2$ dimensions, so we do not discuss the values of these integrals further. 

The first non-trivial differential equation is the one satisfied by $I_0$:
\begin{equation}
\begin{aligned}
	\frac{\mathrm d}{\mathrm dx} I_0 	&=	\frac{1+x^2}{(1-x)(1+x)x} I_0	-	\frac2{(1-x)(1+x)}I_{0,0} + \mathcal O(d-2)	\, ,
\label{eq:gmI0}
\end{aligned}
\end{equation}
which determines the integral corresponding to a one-loop bubble times $(l-1)$ one-loop tadpoles. 
The value of this integral is of course known, and we evaluate it explicitly 
(without recourse to the differential equation~\eqref{eq:gmI0}) in appendix~\ref{app:derivationmaster}. The result is:
\begin{equation}
\begin{aligned}
	I_0	=	-\frac {2x}{1-x^2}\log x + \ord(d-2)\, .
\label{J0full}
\end{aligned}
\end{equation}

The remaining master 
integrals can be organised into two sets $\underline\cI^\pm_l$, where the maximal 
cuts of the first and second set satisfy the homogeneous differential equation of the $(l-1)$-loop banana graph evaluated at $x^{\pm1}$ respectively. 
The non-trivial master integrals in the top sector in $d=2$ dimensions satisfy the equations:
\begin{equation}
\begin{aligned}
	\frac{\mathrm d}{\mathrm dx}\underline{\cI}^+_{l}	&=	\mathbf{GM}_\text{ban}^{(l-1)}(x) \underline{\cI}^+_{l}	+	\underline{N}_l^{+} I_0	+ \mathcal O(d-2)\,,\\
	\frac{\mathrm d}{\mathrm dx} \underline{\cI}^-_{l}	&=	\mathbf{GM}_\text{ban}^{(l-1)}(1/x) \underline{\cI}^-_{l}	+	\underline{ N}_l^{-} I_0 + \mathcal O(d-2)\,,
\label{gmJrest}
\end{aligned}
\end{equation}
with 
\begin{equation}
\begin{aligned}
	\underline{N}_l^{+}	&=	\begin{pmatrix}
							\underbrace{0	\ , \hdots \ , 0}_{l-2} \ ,	\frac{l!}2\frac{(1-x)(1+x)}{x^{l-1}\prod_{p\in\Delta^{(l-1)}}(1-p x)}
						\end{pmatrix}^T		\,,	\\
	\underline{ N}_l^{-}	&=	\begin{pmatrix}
							\underbrace{0	\ , \hdots \ , 0}_{l-2} \ ,	(-1)^l\frac{l!}2\frac{(1-x)(1+x)}{x^{\lfloor l/2\rfloor+1}\prod_{p\in\Delta^{(l-1)}}(p- x)}
						\end{pmatrix}^T			\, .				
\label{gmbanana}
\end{aligned}
\end{equation}
Here $\Delta^{(l)}$ is the set of singularities of the differential equation of the $l$-loop banana integral. It is given by 
$\Delta^{(l)} = \bigcup_{j=0}^{\lceil\frac{l-1}2\rceil}\left\{ (l+1-2j)^2 \right\}$. A detailed and convenient discussion of the 
banana integrals, including an analysis of their differential equations, can be found in refs.~\cite{Bonisch:2020qmm,Bonisch:2021yfw}. 
As a consequence of our basis choice, we find (cf.~eq.~\eqref{eq:3-loop_derivative}):
\beq
\cI_{l,k}^{\pm} = \frac{\rd^{k-1}}{\rd x^{k-1}}\cI_{l,1}^{\pm}\,,\qquad k=2,\hdots, l-1\,.
\eeq
Combined with eq.~\eqref{eq:parity}, we see that all master integrals are determined by $\cI_{l,1}^{+}$. The latter integral can be obtained as a solution to an inhomogeneous $(l-1)^{\textrm{th}}$-order inhomogeneous Picard-Fuchs equation by decoupling the Gauss-Manin system.\footnote{The factorisation of our differential equation compared to~\cite{Lairez:2022zkj} is due to the usage of the Landau variable.} One finds for the $l$-loop case that\footnote{We can write down the same equation for $\cI_{l,1}^-$ under the exchange $x \to 1/x$.}
\begin{equation}
\begin{aligned}
	\mathcal L_{\text{ban},l-1} \cI_{l,1}^+	=	-(-1)^l\,l!/2\,(1-x^2)\,I_0 + \mathcal O(d-2) \, .
\label{decoupled}
\end{aligned}
\end{equation}
The differential operators $\mathcal L_{\text{ban},l-1}$ annihilate the maximal cuts of the $(l-1)$-loop banana integral with unit numerator and propagator powers. It is hard to write down an all-loop form for $\mathcal L_{\text{ban},l-1}$, but there are efficient algorithms to generate the operators to high loop orders~\cite{Vanhove:2014wqa,Bonisch:2020qmm}.
As an example, for low loop orders we have:
\begin{align}
\nonumber	\mathcal L_{\text{ban},1} =& 1-2 x-(1-4 x)\theta \,,	\\[1ex]
\nonumber	\mathcal L_{\text{ban},2} =& 1-3 x-(2-10 x) \theta +(1-x) (1-9 x) \theta ^2\,,	\\[1ex]
\label{listbanop}	\mathcal L_{\text{ban},3} =& 1-4 x-(3-18 x) \theta +(3-30 x) \theta ^2-(1-4 x) (1-16 x) \theta ^3	\,,\\[1ex]
\nonumber	\mathcal L_{\text{ban},4} =& 1-5 x-(4-28 x) \theta +\left(6-63 x+26 x^2-225 x^3\right) \theta ^2-\left(4-70 x+450 x^3\right) \theta ^3 \\
		\nonumber	&\quad+(1-x) (1-9 x) (1-25 x) \theta ^4	\, ,
\end{align}
with $\theta=x\frac{\rd}{\rd x}$.



\section{Solving the Gauss-Manin equation for ice cone integrals}
\label{subsec:mastersiterated}

In this section we show how we can explicitly solve the GM equation~\eqref{gmJrest} for $\underline{\cI}^+_l$ in $d=2$ dimensions. We follow closely the strategy of ref.~\cite{Bonisch:2021yfw}, which we briefly recall here for convenience. Let us define the following fundamental solution matrix:
\begin{equation}
	\bfW_{l-1}^+(x)	=	\begin{pmatrix}
							\Pi_{l-1,1}(x)	&	\hdots	&	\Pi_{l-1,l-1}(x)	\\
							\partial_x\Pi_{l-1,1}(x)	&	\hdots	&	\partial_x\Pi_{l-1,l-1}(x)	\\
							\vdots	&	\hdots	&		\vdots		\\
							\partial_x^{l-2}\Pi_{l-1,1}(x)		&	\hdots	&	\partial_x^{l-2}\Pi_{l-1,l-1}(x)
						\end{pmatrix}	\,,
\label{cutsban}
\end{equation} 
which satisfies the homogeneous version of eq.~\eqref{gmJrest}:
\beq
\frac{\rd}{\rd x}\bfW_{l-1}^+(x) = \mathbf{GM}_\text{ban}^{(l-1)}(x) \bfW_{l-1}^+(x)\,.
\eeq
Since $ \mathbf{GM}_\text{ban}^{(l-1)}(x)$ describes the GM equation satisfied by the $(l-1)$-loop banana integrals, the fundamental solution matrix $\bfW_{l-1}^+(x)$ coincides with the one for the latter integrals. Consequently, the entries of $\bfW_{l-1}^+(x)$ describe the periods $\Pi_{l,k}(x)$ of the one-parameter family of Calabi-Yau $(l-2)$-folds attached to the $(l-1)$-loop banana integrals
(cf.,~e.g.,~ref.~\cite{Bonisch:2021yfw} for a precise definition of these functions). In particular, they are algebraic for $l=2$, and they can be expressed in terms fo elliptic integrals of the first kind for $l=3$ and $l=4$. Starting from five loops, it is not known how to express the periods in terms of known functions, and it is expected that this is not possible. Nevertheless, periods of Calabi-Yau $(l-2)$-folds
have properties which allow to evaluate them efficiently, cf. ref.~\cite{Bonisch:2020qmm}.\footnote{The code is available from \href{http://www.th.physik.uni-bonn.de/Groups/Klemm/data.php}{http://www.th.physik.uni-bonn.de/Groups/Klemm/data.php} and in the supplementary material of ref.~\cite{Bonisch:2020qmm}.} In particular, it is known that the GM system has a point of maximal unipotent monodromy (MUM)  at $x=0$. We choose the periods to form a Frobenius basis close to this MUM point, such that
\beq\label{eq:Pi_norm}
\Pi_{l-1,k}(x) \sim \frac{x}{(k-1)!}\,\log^{k-1}x + \ord(x^2) \textrm{~~as~~} x\to0\,.
\eeq
We also define the $l\times l$ skew-diagonal matrix:
\begin{equation}
\begin{aligned}
	{\bf\Sigma}_l	&=	\left(\begin{smallmatrix}
      						   & & & 1  \\
      						   & & -1 & \\
      						   & 1 & & \\
      						   \iddots & & &
    					\end{smallmatrix}\right)		\, ,
\label{intersection}
\end{aligned}
\end{equation}
which is symmetric for $l$ odd and symplectic for $l$ even. 
From  Griffiths transversality conditions of the Calabi-Yau manifold associated to the banana graph, one can derive quadratic relations between the Frobenius basis elements, and so also for the maximal cuts of the banana graph~\cite{Bonisch:2021yfw}. These quadratic relations can conveniently be written as a matrix relation
\begin{equation}
	\textbf{Z}_l^+	=	\textbf{W}_l^+ {\bf \Sigma}_l \left(\textbf{W}_l^+\right)^{T}\,,
	\label{quadrelations}
\end{equation}
where $\textbf{Z}_l^+$ is an explicitly computable matrix of rational functions~\cite{Bonisch:2021yfw}.
From this equation we can extract an explicit representation for the inverse of the fundamental solution matrix:
\begin{equation}
\left(\textbf{W}_l^{+}\right)^{-1}	=	{\bf \Sigma}_l\left(\textbf{W}_l^+\right)^{T}\left(\textbf Z_l^+\right)^{-1}\, .
\end{equation}
For more details about these quadratic relations and the following considerations we refer to ref.~\cite{Bonisch:2021yfw}.

With this setup, we can now easily solve the GM equation~\eqref{gmJrest} for $\underline{\cI}_l^+$ following the recipe of ref.~\cite{Bonisch:2021yfw}. We find:
\begin{equation}
\begin{aligned}
	\underline{\cI}_l^+(x;2)	&=	\textbf{W}_{l-1}^+(x) {\bf \Sigma}_{l-1}\int_{\Vec{1}_0}^x {\textbf{W}}_{l-1}^{+T}(\xh)\textbf{Z}_{l-1}^+(\xh)^{-1} \underline{N}_B^{(l-1)}(\xh)\,    I_0(\xh)	\, \mathrm d\xh + \textbf{W}_{l-1}^+(x)\underline c^+_l	\\
							&=	-(-1)^ll!\,\textbf{W}_{l-1}^+(x){\bf \Sigma}_{l-1} \int_{\Vec{1}_0}^x	\frac{\underline{\Pi}_{l-1}(x)}{\xh^2} \,\log (\xh) \mathrm d\xh	+ \textbf{W}_{l-1}^+(x)\underline c^+_l	\,,
\label{mastersit}
\end{aligned}
\end{equation}
where $\underline c^+_l=(c^+_{l,1},\hdots,c^+_{l,l-1})$ is a constant vector and we introduced the notation $\underline{\Pi}_l = (\Pi_{l,1},\ldots,\Pi_{l,l})^T$.
The lower integration boundary chosen as $\Vec{1}_0$ indicates that we use tangential base point regularization with base point $\Vec{1}_0$ to deal with the divergent integrals in eq.~\eqref{mastersit}. For more details we refer to refs.~\cite{Brown:mmv,Bonisch:2021yfw}. We can extract the following expression for the master integral $\cI_{l,1}^+$:
\beq\label{eq:I+l1_result}
\cI_{l,1}^+(x;2) = \sum_{k=1}^{l-1}\Pi_{l-1,l-k}(x)\,\left[c^+_{l,l-k} + l!\,(-1)^{k} I(f_{l-1,k},f_{0};x)\right]\,,
\eeq
where we introduced the iterated integral~\cite{ChenSymbol}:
\beq\label{eq:II_def}
I(f_1,\ldots,f_n;x)  = \int_{\vec{1}_0}^{x}\rd x'\,f_1(x')\,I(f_2,\ldots,f_n;x') \,,\qquad I(;x)=1\,.
\eeq
The integration kernels appearing in our result are
\beq
f_{0}(x) = \frac{1}{x}\textrm{~~~and~~~} f_{l,k}(x) = \frac{1}{x^2}\,\Pi_{l,k}(x) = \frac{1}{x\,(k-1)!}\log^{k-1}x + \ord(x^0)\,.
\eeq
Equation~\eqref{eq:I+l1_result} is one of the main results of this paper, and it expresses $\cI_{l,1}^+(x;2)$ as a linear combination of double iterated integrals involving the periods of the Calabi-Yau variety of the $(l-1)$-loop banana integral (the integration constants $\underline c^+_l$ will be fixed shortly). The result is strikingly similar to the corresponding result for the $(l-1)$-loop banana integral in $d=2$ dimensions with unit numerator and propagator powers of ref.~\cite{Bonisch:2021yfw}, which we reproduce here for convenience:
\beq\label{eq:ban_result}
\textrm{Ban}_{l-1,1}(x;2) = -(-1)^ll!\sum_{k=1}^{l-1}\Pi_{l-1,l-k}(x)\,(-1)^k\,\left[\frac{\lambda^{(k)}_0}{(k+1)!} 
+  I(f_{l-1,k};x)\right]\,.
\eeq
We see that the results for $\frak{l}=(l-1)$-loop banana and $l$-loop ice cone integrals are almost identical, but the ice cone integrals involve double iterated integrals. In the case of the banana integrals, it is possible to obtain the integration
constants $\lambda^{(k)}_0/(k+1)!$, which are related to the contribution of the unique holomorphic period in the Feynman integral (called $\lambda^{(\frak{l})}_0$ in ref.~\cite{Bonisch:2020qmm}), from a generating series:
\beq
\sum_{\frak{l}=0}^{\infty}\frac{\lambda^{({\frak l})}_{0}}{(\frak{l}+1)!}\,{t^\frak{l}} 
= -\frac{\Gamma(1-t)}{\Gamma(1+t)}\,e^{-2\gamma t-i\pi t}\,,\qquad \gamma =-\Gamma'(1)\, .
\label{eq:leading} 
\eeq
For the case of the Banana integral one gets in addition the contributions of periods with 
leading behaviour $\sim\log(x)^k$ in the Feynman integral by
\beq
\lambda^{(\frak{l})}_k=(-1)^k\left(\frak{l}+1\atop k\right)\lambda^{(\frak{l}-k)}_k \ .
\label{eq:subleading} 
\eeq
Both informations in eqs.~\eqref{eq:leading} and \eqref{eq:subleading} were  conjectured in 
ref.~\cite{Bonisch:2020qmm} to arise from  a novel $\widehat \Gamma$-class evaluated in $(\mathbb{P}^1)^{\frak{l}+1}$.  
This  allows one to evaluate the corresponding  values also in the non-equal mass case~\cite{Bonisch:2020qmm}, and the 
claim was later proven in ref.~\cite{Iritani:2020qyh}.\footnote{It can be  
extended to the $\epsilon$ deformed case using a Barnes integral representation for the banana graph integral~\cite{Bonisch:2021yfw}. Such extensions of the formalism of $\widehat \Gamma$-classes  over the dimensions 
are called motivic $\widehat \Gamma$-class formalisms in mathematics.} In the next section we argue that the integration constants $\underline c^+_l$ can be obtained from a 
very similar structure.

\subsection{Boundary conditions and analytic structure for ice cone graphs}
\label{subsec:lincom}  

The goal of this section is to fix all the boundary conditions $\underline c^+_l$ in eq.~\eqref{eq:I+l1_result} needed to obtain the final result for $\cI_{l,1}^+$. We do this by studying the structure of the differential equation satisfied by $\cI_{l,1}^+$ to infer how it behaves close to singular points. This is most conveniently done by understanding 
$l$-loop ice cone integrals as a double extension of the maximal cut of $l-1$ banana integral.
Due to the special form of the master integral $I_0$, it turns out that we can homogenize the differential equation \eqref{decoupled} simply by acting with $(\theta-1)^2$ from the left. This means that the ice cone integral $\cI_{l,1}^+$ in two dimensions satisfies the double extended differential equation of the maximal cuts of the $(l-1)$-loop banana integral, i.e.
\begin{equation}
\begin{aligned}
	\mathcal L_{\text{ice},l} \cI_{l,1}^+	=	(\theta-1)^2\mathcal L_{\text{ban},l-1} \cI_{l,1}^+	=	\ord(d-2)\, .
\label{doubext}
\end{aligned}
\end{equation}
In ref.~\cite{Bonisch:2020qmm,Klemm:2019dbm} it was shown that the full $(l-1)$-loop banana integral is annihilated by the first extension, i.e., by $(\theta-1)\mathcal L_{\text{ban},l-1}$. The operator $\mathcal L_{\text{ice},l}$ has a MUM point at $x=0$, so that all $l+1$ indicials at that point are equal to one. 

\begin{figure}[!thb]
    \centering
    \begin{tikzpicture}
    \draw (12.6,2.2) node[label=below:{$x$-plane}] {};
    \draw [thick] (11.8,1.5)--(13.3,1.5);
    \draw [thick] (11.8,1.5)--(11.8,1.9);
    \draw [-stealth] (1,0)--(12.5,0);
    \draw (12.5,0) node[label=above:{{\footnotesize{Re($x$)}}}] {};
    \draw [-stealth] (6,-0.2)--(6,0.9);
    \draw (6,0.9) node[label=right:{{\footnotesize{Im($x$)}}}] {};
    \draw (2.2,0) node[circle,fill,inner sep=1.5pt,label=below:{\hspace{5pt}$x=-\infty$}] {};
    \draw (2.5,-1.5) node[label=below:{sing. point}] {};
    \draw (2.5,-.75) node[label=below:{sing. point}] {};
    \draw (6,0) node[circle,fill,inner sep=1.5pt,label=below:{$x=0$}] {};
    \draw (6,-1.5) node[label=below:{MUM point}] {};
    \draw (6,-.75) node[label=below:{MUM point}] {};
    \draw (8.5,0) node[circle,fill,inner sep=1.5pt,label=below:{\hspace{10pt}$x=\frac{1}{l^2}$}] {};
    \draw (8.5,-1.5) node[label=below:{sing. point}] {};
    \draw (8.5,-.75) node[label=below:{reg. point}] {};
    \draw (10.7,0) node[circle,fill,inner sep=1.5pt,label=below:{\hspace{10pt}$x=\frac{1}{(l-2)^2}$}] {};
    \draw (10.7,-1.5) node[label=below:{sing. point}] {};
    \draw (10.7,-.75) node[label=below:{reg. point}] {};
    \draw (11.7,0) node {{\huge$\cdots$}};
    \draw (-0,-1.5) node[label=below:{diff. eq.:}] {};
    \draw (-0,-.75) node[label=below:{$\cI^{+}_{l,1}$:}] {};
    \draw [very thick,snake it] (2.2,0)--(6.06,0);
    \end{tikzpicture}
    \caption{Global analytic structure of the master integral $\cI^{+}_{l,1}$ compared to the singularities of the differential operator $\mathcal L_{\text{ice},l}$. $\cI^{+}_{l,1}$ has a branch cut from $x=0$ to $x=-\infty$.}
    \label{fig:globalana}
\end{figure}
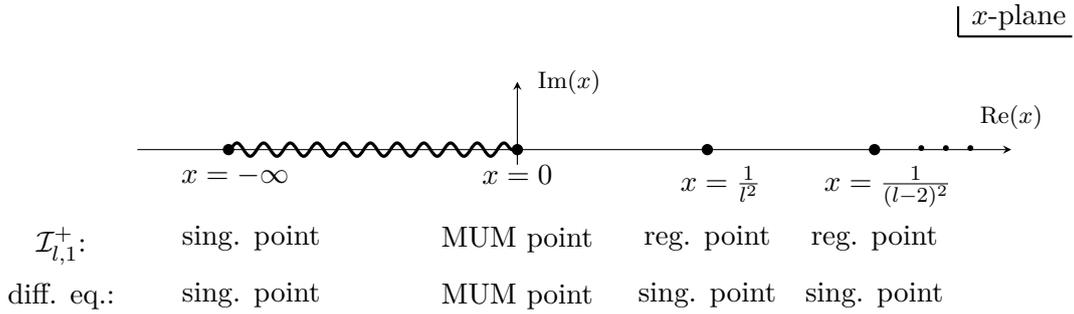

Therefore, it makes sense to consider a Frobenius basis around the MUM point which is, of course, related to the basis in eq.~\eqref{mastersit} given in terms of iterated integrals of Calabi-Yau periods. 
%
The coefficients $\underline c_l^+$ in eq.~\eqref{eq:I+l1_result} can be determined from regularity properties of the integral $\cI_{l,1}^+$. The integral $\cI_{l,1}^+$ has a branch cut starting at $x=0$ and ending at $x=-\infty$. Around all the other singular points of the differential equation \eqref{doubext} the integral is regular, see figure \ref{fig:globalana}. In other words, the only non-trivial monodromies of the integral $\cI_{l,1}^+$ are at $x=0,-\infty$. This determines the coefficients $\underline c_l^+$ for loop orders $l=2,3$ completely. For $l=4,5$ we also used numerical evaluations of the integral to fix all coefficients. For this we numerically integrated the Feynman parameter representation for various values of the Landau variable $x$. Here we used the first $l-1$ values of $x$ to fix the parameters $\underline c_l^+$ in eq.~\eqref{eq:I+l1_result} and further values of $x$ to check agreement with our result in eq.~\eqref{eq:I+l1_result}. Our findings are listed in table \ref{tab:coeffs0}.
\begin{table}[!th]
{\footnotesize
\begin{center}\begin{tabular}{c?cccc}
	$l $    &$ c_{l,1}^+      $&$ c_{l,2}^+ $&$c_{l,3}^+$&$ c_{l,4}^+ $ \\[1ex]\toprule[1.5pt]
	$2$   &$-2\zeta(2)     $&$              $&$             $&$                $ \\ [ 1mm]
         $3$   &$ 4  \zeta(3)   $&$  - 6 \zeta(2)           $&$              $&$                $ \\ [ 1mm]
         $4$   &$-42 \zeta (4) $&$      16 \zeta(3)$&$ -24 \zeta(2)            $&$                $ \\ [ 1mm]
         $5$   &$ 48\zeta(5) + 80\zeta(2)\zeta(3)             $&$   -210\zeta(4)           $&$     80\zeta(3)             $&$  -120 \zeta(2)              $
\end{tabular}
\end{center}}
\caption{Numerically determined coefficients $ c_{l,k}^+$ for the master integral $\cI^+_{l,1}$ for $2\le l \le 5$.}
\label{tab:coeffs0}
\end{table}
From these data we found some interesting patterns: First, we observe that the $c_{l,k}^+$ are polynomials in zeta values of uniform weight $l-k+1$. Second, the coefficients $c_{l+1,k+1}^+$ are obtained from $c_{l,k}^+$ for $1\le k \le l-1$ by multiplication with $l+1$:
\beq
c_{l+1,k+1}^+ = (l+1)\,c_{l,k}^+\,.
\label{crelations}
\eeq
Finally, one can write down a generating series for the coefficients $c_{l,1}^+$ for different loop orders:
\begin{equation}
\begin{aligned}
	1 +	\sum_{l=2}^\infty (-1)^{l+1}c_{l,1}^{+} \frac{t^l}{l!} 	=	\Gamma(1-t)^2 e^{-2\gamma t}	\, .
\label{genseries}
\end{aligned}
\end{equation}
With both relations we can conjecturally\footnote{For $l=6,7$ we have also numerically checked our result in eq.~\eqref{eq:I+l1_result} with the coefficients $\underline c^+_l$ from eqs.~\eqref{crelations} and \eqref{genseries} and found good agreement.} determine for arbitrary loop order $l$ the coefficients $\underline c_l^+$ in eq.~\eqref{eq:I+l1_result}. Note that eqs.~\eqref{genseries} and \eqref{crelations} are analogous to 
eqs.~\eqref{eq:subleading} and \eqref{eq:leading} for the banana integral, which can be viewed as a single 
extension of its maximal cut integral. Moreover the dependence of the transcendentality of 
the coefficients on the corresponding loop order $l$ and $\frak{l}$ is the same. We expect that a 
$\widehat \Gamma$, or even a motivic $\widehat \Gamma$, formalism exists for the ice cone graph as well. 
However, the notion of `geometry' in which these $\widehat \Gamma$-classes have to be evaluated  has probably 
to be considerably generalised.

The word \emph{extension} for studying a differential operator ${\cal L}^{(2)} {\cal L}^{(1)}$  
and its solution space in comparison with the differential  operator ${\cal L}^{(1)}$  and its 
solution space seems very high browsed. However, clearly the solution space of the latter is a subspace  
of the former, and since both of them are representations of the Galois group, the word extension 
is used here exactly in the same way as in the context of groups and their representations. So our simple 
observation that ice cone integrals are double extentions of maximal cut banana integrals, with the 
full banana integral in between, yields a very interesting mathematical structure that could lead to a proof of  
eqs.~\eqref{genseries}, \eqref{crelations} from eqs.~\eqref{eq:subleading}, \eqref{eq:leading}. The 
structure of the extension for the ice cone graph will be further discussed in section \ref{sec:extensions}.

%% file: structure.tex
\section{A canonical choice of coordinate}
\label{sec:structure series}

\subsection{The structure series and the mirror map}
In the previous section we have shown that it is possible to write the ice cone integrals to arbitrary loop order as iterated integrals involving the periods of a Calabi-Yau variety, similarly to the result obtained for the banana integrals in ref.~\cite{Bonisch:2021yfw}. The integration variable $x$ parametrises the moduli space of the family of Calabi-Yau $n$-folds associated to the banana integrals (where $n=l-1$ or $l-2$ for banana and ice cone integrals respectively). This choice of coordinate is not canonical. A canonical choice of the coordinate on the moduli space in a neighbourhood of the MUM point $x=0$ is
\beq\label{eq:t_def}
t(x) = \frac{1}{2\pi i}\frac{\Pi_{n+1,1}(x)}{\Pi_{n+1,2}(x)} = \frac{1}{2\pi i}\log x + \ldots\,,
\eeq
where the dots indicate terms that are analytic in a neighbourhood of $x=0$. The inverse to eq.~\eqref{eq:t_def} is the \emph{mirror map}:
\beq
x(q) = q + \ord(q^2)\,,\qquad q = e^{2\pi i t}\,.
\eeq
Using this canonical coordinate, the homogeneous Picard-Fuchs operator of a one-parameter family of Calabi-Yau $n$-folds has a characteristic form. To see this, we start with the operator $\cL_{n} = \sum_{k=0}^{n+1}A_k(x)\theta^k$ and write it in the following form~\cite{bogner2013algebraic,Bognerthesis}:
\beq
\cL_n = \beta(x)\,\theta\alpha_n(x)\theta\alpha_{n-1}(x)\theta\cdots\theta\alpha_1(x)\theta\alpha_0(x)\,,
\eeq
where $\beta(x)$ fixes the normalisation and the sequence $\alpha_0(x),\ldots,\alpha_n(x)$ is called the \emph{structure series} of the family of Calabi-Yau $n$-folds. The $\alpha_m(x)$ are holomorphic in a neighbourhood of the MUM point $x=0$. They can be computed as $\alpha_m(x) = u_{m,m}(x)^{-1}$, where the $u_{m,k}(x)$ are determined recursively:
\beq\label{eq:u_rec}
u_{m,k}(x) = \theta\left[\frac{u_{m-1,k}(x)}{u_{m-1,m-1}(x)}\right] \textrm{~~~and~~~} u_{0,k}(x) = \Pi_{n+1,k+1}(x)\,.
\eeq
They satisfy the symmetry property $\alpha_k(x) = c\,\alpha_{n-k+1}(x)$ for some non-zero $c\in\mathbb{C}$. In particular, we have $\alpha_0(x) = \Pi_{n+1,1}(x)^{-1}$ and $\alpha_1(x) = [2\pi i\,\theta t(x)]^{-1}$. It follows that $\cL_{n}$ can be written as a differential operator in $q$:
\beq
\cL_n = C\,\beta(x)\,\widehat\alpha_1({q})^{-1}\,\widehat{\cL}_{n,q}\,\widehat{\Pi}_{n+1,1}(q)^{-1}\,,
\eeq
where $C$ is a complex constant, $\widehat\alpha_m(q) = \alpha_m(x(q))$ and $\widehat{\Pi}_{n+1,k}(q) = {\Pi}_{n+1,k}(x(q))$. $\widehat{\cL}_{n,q}$ is the normalised operator~\cite{bogner2013algebraic,Bognerthesis}:
\beq\label{eq:Lhat_q}
\widehat{\cL}_{n,q} = \theta_q^2\frac{1}{Y_{n,1}(q)}\theta_q\frac{1}{Y_{n,2}(q)}\theta_q \cdots\theta_q\frac{1}{Y_{n,2}(q)}\theta_q\frac{1}{Y_{n,1}(q)}\theta_q^2\,,
\qquad \theta_q = q\frac{\rd}{\rd q}\,.
\eeq
The functions 
\beq
\label{Yinvariants}
Y_{n,k}(q) = \frac{\widehat\alpha_{1}(q)}{\widehat\alpha_{k+1}(q)}\,,  \quad \text{for } 1\le k\le n-2\,,
\eeq
are holomorphic in a neighbourhood of $q=0$ and are called the \emph{$Y$-invariants} of the differential operator $\cL_n$~\cite{bogner2013algebraic}.

For small values of $n$ we list these differential operators here:
\beq\bsp
\label{eq:oneparameterform}
\widehat\cL_{1,q} &\,=\theta_q^2\,,\\
\widehat\cL_{2,q} &\,=\theta_q^3\,,\\
\widehat\cL_{3,q} &\,=\theta_q^2\frac{1}{Y_{3,1}}\theta_q^2\,,\\
\widehat\cL_{4,q} &\,=\theta_q^2\frac{1}{Y_{4,1}}\theta_q\frac{1}{Y_{4,1}}\theta_q^2\,,\\
\widehat\cL_{5,q} &\,=\theta_q^2\frac{1}{Y_{5,1}}\theta_q\frac{1}{Y_{5,2}}\theta_q\frac{1}{Y_{5,1}}\theta_q^2\,.
\esp\eeq
Through $n=4$, this agrees with the Picard-Fuchs operators for the banana graphs given in refs.~\cite{Pogel:2022ken,xing_private}. Starting from $n=5$, we see that we need two different $Y$-invariants.

We see that the differential operators are entirely determined by the $Y$-invariants. For later convenience, we define $Y_{n,0}(q) = Y_{n,n-1}(q)=1$. Since these quantities play such an important role, we list here some of their properties~\cite{Greene:1993vm,bogner2013algebraic}.  First of all, the $Y$-invariants satisfy the reflection identity
\beq\label{eq:Y_ref}
Y_{n,k}(q) = Y_{n,n-1-k}(q)\,.
\eeq
Second, the $Y$-invariants are not independent, but they are related by the identity:
\beq
\label{relationYinva}
\prod_{k=1}^{n-2}Y_{n,k}(q) = \left(\frac{x(q)}{\widehat\Pi_{n+1,1}(q)}\right)^2\frac{\widehat\alpha_1({q})^n}{\prod_{p\in\Delta^{(n+1)}}(1-px(q))}  \,.
\eeq
This was also noticed for Calabi-Yau three-folds in refs.~\cite{Pogel:2022ken}.

\subsubsection{The string theory and algebraic geometry perspective}
Let us connect the discussion of the previous section to the information available from 
the mathematical string theory literature on the structure of Picard-Fuchs ideals, or equivalently of the Gauss-Manin connections, for Calabi-Yau $n$-folds $X$ in  {\sl inhomogeneous flat coordinates}, such as in eq.~\eqref{eq:t_def}. 
The $Y_{n,k}$ have been introduced in ref.~\cite{Greene:1993vm} (there called $Y^1_k$), together with a general algorithm to calculate them from the 
Gauss-Manin system. They  were explicitly evaluated 
for the one-parameter mirrors of Calabi-Yau $n$-folds for $n=4,\ldots,10$, defined as degree $n+2$ hypersurfaces 
in $\mathbb{P}^{n+1}$ with a hypergeometric Picard-Fuchs operator  of the form 
${\cal L}_n=\theta^{n+1}-(n+2)x\prod_{k=1}^{n+1} [(n+2)\theta+k]$. Their general properties in eqs.~\eqref{eq:Y_ref} and \eqref{relationYinva} and the number of independent $Y_{n,k}$ have also been pointed out in ref.~\cite{Greene:1993vm} in a formalism 
that generalises to the multi-moduli case. 

To understand the latter, let us recall (see refs.~\cite{MR3965409,Bonisch:2021yfw} for reviews), that Griffiths duality in eq.~\eqref{eq:griffith} implies in the Calabi-Yau three-fold case the existence of a  
prepotential $F(\ux)=\frac{1}{2} (X^I F_I)$, where ${\underline{\omega}}^T=(F_I,X^I)$, $I=0,\ldots, h_{n-1,1}(X)$ 
is the period vector in any symplectic basis. Griffiths transversality also implies that $F$ is homogeneous 
of degree two in the {\sl local homogenoeus coordinates} $X^I$, and $F_I=\partial F/\partial X^I$ is homogeneous of degree one in $X^I$. Moreover in {\sl inhomogeneous flat coordinates} 
$t^i=X^i/X_0$, $i=1,\ldots, h_{n-1,1}(X)$, that can be chosen at any point in the moduli space with $X^0(\ux)\neq 0$, 
it implies that we can introduce an inhomogeneous prepotential ${\cal F}(\ut) =F(\ut)/(X^0(\ut))^2$ so that the multiparameter generalisation\footnote{So that $Y_{3,1}=C_{111}$.}  of $Y_{3,1}$, called $C_{ijk}$, is  given by $C_{ijk}=\partial_{t_i}\partial_{t_j} \partial_{t_k}{\cal F}(\ut)$. We can conclude 
from the above facts immediatley that the period vector can be written as 
${\underline{\omega}}=X^0 (2 {\cal F}-t^i\partial {\cal F},\partial_i {\cal F}, 1,t^i)^T$. By change  
of the dependent variable,\footnote{Which is referred to as weight zero K\"ahler 
gauge in algebraic geometry.} one  defines a vector $\underline{\cal V}=
(2 {\cal F}-t^c\partial_c {\cal F}, \partial_j (2 {\cal F}-t^c\partial_c {\cal F}),t^j,1)^T$, and 
with ${\cal V}^j\coloneqq{\cal V}_{b_3(X)/2+j}$, ${\cal V}^0\coloneqq{\cal V}_{b_3(X)}$ one gets  
\begin{equation}
\label{eq:flat}
\partial_{t^i} \left(\begin{array}{c} {\cal V}_0\\ {\cal V}_j\\  {\cal V}^j\\{\cal V}^0\end{array}\right) =
\left(\begin{array}{cccc} 
0&\delta_{ik}&{ 0} &0\\ 
0&{0}&C_{ijk} &0\\
0&{0}&{0}& \delta_{i}^j\\ 
0&{0}&{0} &0\end{array} 
\right)\left(\begin{array}{c}  {\cal V}_0\\ {\cal V}_k\\ {\cal V}^k \\{\cal V}^0\end{array}\right)\ .
\end{equation}
This is the Gauss-Manin connection in projective flat coordinates and  $q_i\frac{\mathrm d}{\mathrm dq_i}=\partial_{t^i}$.  Hence writing eq.~\eqref{eq:flat} in the fourth-order form in the one-parameter specialisation  yields  the third 
equation in eq.~\eqref{eq:oneparameterform}.  Note, however, that eq.~\eqref{eq:flat}  holds in any 
flat inhomogeneous coordinates. 

The generalisation of eq.~\eqref{eq:flat} to multiparameter 
$n$-folds appeared for the first time in ref.~\cite{CaboBizet:2014ovf} (see the review in ref.~\cite{MR3965409}). It has 
also been pointed out there (see also refs.~\cite{Klemm:1996ts,Mayr:1996sh} for the Calabi-Yau 
four-fold multi-parameter cases) that the three-point functions $C_{ijk}$, as well as their generalisations  
to the three-point functions of multiparameter Calabi-Yau $n$-folds, can be identified 
with three-point functions in the topological sector of a  $(2,2)$ superconformal 2d 
field theory, and as such fulfill a {\sl graded Frobenius algebra}. Geometrically, the latter 
can likewise be derived as a consequence of Griffiths transversality, and for example eq.~\eqref{relationYinva} is a consequence of the well-known factorisation of the $n$-point function 
into three-point functions in a 2d CFT. The grading comes from the $U(1)$ charges 
under the two $U(1)$ symmetries in the two copies of the $N=2$ superconformal algebras~\cite{Lerche:1989uy}.
There are two twists to obtain topological sectors: The $A$-twist projects to the diagonal $U(1)$
and all fields in this topological sector have charges $(q,q)$. The B-twist projects to the anti-diagonal $U(1)$ and all fields in this topological sector have charges $(q,-q)$. Mirror symmetry
exchanges these topological theories. At the  MUM point this charge grading can be identified with
the logarithmic degenerations of the periods, but it is globally defined in the deformation 
space.  Specialising the generalisations of eq.~\eqref{eq:flat} in  refs.~\cite{CaboBizet:2014ovf,MR3965409} to the $(n+1)^{\textrm{th}}$-order form in the one-parameter case yields all equations in eq.~\eqref{eq:oneparameterform}. 

At the MUM point and expanded in the $q$-coordinates, the coefficients of the three-point functions count Gromov-Witten invariants of holomorphic maps to  the mirror $\check X$ to $X$. The mirror $\check X$ 
is given for the particular case of the banana integrals in refs.~\cite{Bonisch:2020qmm,Bonisch:2021yfw}. The Gromov-Witten invariants have multi-covering  formulas relating them to {\sl integer geometric invariants} 
of holomorphic curves embedded into $\check X$. These  multi-covering formulas and the definition of the corresponding  
invariants  have been found in ref.~\cite{Aspinwall:1991ce} for the three-fold case, in ref.~\cite{MR2415462}  
for the four-fold case and in ref.~\cite{MR2683212} for the five-fold case.

\subsubsection{Calabi-Yau operators for banana and ice cone integrals}
Before we continue, let us give some explicit results for the family of Calabi-Yau varieties which appear as the maximal cut geometries in the banana and ice cone graphs. Let us start with the elliptic case, which is relevant for the two-loop banana integral as well as three-loop ice cone integral. The mirror map is given by
\beq\label{eq:mirror_map_n=1}
    x(q)    =  \left(\frac{\eta(t)\eta(6t)^2}{\eta(2t)^2\eta(3t)}\right)^4 = q - 4q^2+10q^3-20q^4+39q^5+\mathcal O(q^6)\,  , 
\eeq
where $\eta(t)$ is the Dedekind eta function.
Note that the mirror map in eq.~\eqref{eq:mirror_map_n=1} is simply a Hauptmodul for $\Gamma_1(6)$~\cite{Maier,Bloch:2013tra}. The holomorphic period  is given by
\begin{equation}
\begin{aligned}
    \widehat{\Pi}_{2,1}(q)    &= \frac{\eta(t)\eta(6t)^6}{\eta(2t)^2\eta(3t)^3} =q-q^2+q^3+q^4+\mathcal O(q^6)  \, ,
\end{aligned}
\end{equation}
which is also related by eq.~\eqref{relationYinva} to the structure series
\begin{equation}
\begin{aligned}
    \alpha_1(x)    &=  (1-x)(1-9x)\left(\frac{{\Pi}_{2,1}(x)}{x}\right)^2  =   1-4x-12x^2-60x^3-348x^4+\mathcal O(x^5)  \,. 
\end{aligned}
\end{equation} 

For the K3 case we have the following mirror map
\beq
    x(q)    =  \left(\frac{\eta(t)\eta(3t)\eta(4t)\eta(12t)}{\eta(2t)^2\eta(6t)^2}\right)^6 = q - 6q^2+21q^3-68q^4+198q^5+\mathcal O(q^6)\,,
\eeq
which is a Hauptmodul for $\Gamma_0(6)^{+3}$~\cite{verrill1996}. The holomorphic period is given by:
\begin{equation}
\begin{aligned}
    \widehat{\Pi}_{3,1}(q)  &=  \left(\frac{\eta(t)\eta(3t)\eta(4t)\eta(12t)}{\eta(2t)\eta(6t)}\right)^2    =  q-2q^2+q^3-4q^4 +\mathcal O(q^6)  \, , 
\end{aligned}
\end{equation}
which is again related by eq.~\eqref{relationYinva} to the structure series
\begin{equation}
\begin{aligned}
    \alpha_1(x)     &= \alpha_2(x)    =  \sqrt{(1-4x)(1-16x)}\,\frac{{\Pi}_{3,1}(x)}{x}   \\
                    &=   1-6x-30x^2-276x^3-3030x^4+\mathcal O(x^5)  \,.  
\end{aligned}
\end{equation}

For the Calabi-Yau three-fold, the mirror map is
\beq
    x(q)    =  q - 8q^2+36q^3-168q^4+514q^5+\mathcal O(q^6) \, .   
\eeq
The only remaining independent invariants can be chosen as the holomorphic period ${\widehat\Pi}_{4,1}(q)$ and the $Y$-invariant
\begin{equation}
\begin{aligned}
    \widehat{\Pi}_{4,1}(q)     &=  q-3q^2+q^3-23q^4-101q^5+\mathcal O(q^6)  \, ,\\
    Y_{3,1}(q)          &=  1+q +17q^2+253q^3+3345q^4+\mathcal O(q^5) \,.   
\end{aligned}
\end{equation}
The structure series follow then from the properties given above.

The Calabi-Yau four-fold case has the same number of independent invariants as the three-fold:
\begin{equation}
\begin{aligned}
    x(q)            &=  q - 10q^2+55q^3-340q^4+955q^5+\mathcal O(q^6) \, ,  \\ 
    \widehat{\Pi}_{5,1}(q)     &= q-4q^2+q^3-64q^4-569q^5+\mathcal O(q^6)  \, ,\\
    Y_{4,1}(q)          &=  1+2q +46q^2+1010q^3+21550q^4+\mathcal O(q^5) \,.   
\end{aligned}
\end{equation}

For the case of the Calabi-Yau five-fold, we get for the first time two independent $Y$-invariants:
\begin{equation}
\begin{aligned}
    x(q)            &=  q - 12q^2+78q^3-604q^4+1425q^5+\mathcal O(q^6) \, ,  \\ 
     \widehat{\Pi}_{6,1}(q)     &=  q-5q^2+q^3-135q^4-1774q^5+\mathcal O(q^6)  \, ,\\
    Y_{5,1}(q)          &=  1+3q +87q^2+2523q^3+74247q^4+\mathcal O(q^5) \,,    \\  
    Y_{5,2}(q)          &=  1+4q +124q^2+3892q^3+123564q^4+\mathcal O(q^5) \,.  
\end{aligned}
\end{equation}

We note that in all cases the coefficients appearing in the $q$-expansions of the mirror map $x(q)$, the holomorphic period $\widehat\Pi_{n+1,1}(q)$ and the $Y$-invariants $Y_{n,m}(q)$ are integers. This is a general feature of Calabi-Yau operators, cf., e.g., refs.~\cite{bogner2013algebraic,Bognerthesis,MR3822913}.

\subsection{$Y$-invariants and iterated integrals}
\subsubsection{Calabi-Yau periods as iterated integrals}
We now show that we can use the $Y$-invariants to write the periods of a Calabi-Yau $n$-fold $\Pi_{n+1,k}(x)$ for $k=2,\hdots,n+1$ (which are not holomorphic at the MUM-point) as iterated integrals involving the $Y$-invariants. To start, let us define the normalised periods
\beq\label{eq:omega_hat_norm}
\widehat{\omega}_{n+1,k}(q) = \frac{\widehat\Pi_{n+1,k}(q)}{\widehat\Pi_{n+1,1}(q)} = \frac{1}{(k-1)!}\,\log^{k-1}q + \ord(q)   \quad\text{for } k=1,\hdots, n+1\,,
\eeq
where the second equality follows from our normalisation of the periods, cf.~eq.~\eqref{eq:Pi_norm}. We define the differential operators
\beq\bsp
\widetilde{\cL}_{0,q}&\, = \theta_q \,,\\
 \widetilde{\cL}_{k,q} &\,= \theta_q\frac{1}{Y_{n,k-1}}\widetilde{\cL}_{k-1,q}\\
 &\, = \theta_q\frac{1}{Y_{n,k-1}}\theta_q\frac{1}{Y_{n,k-2}}\theta_q\cdots \theta_q\frac{1}{Y_{n,1}}\theta_q\frac{1}{Y_{n,0}}\theta_q     \quad\text{for } k=1,\hdots,n\,.
\esp\eeq
Note that eq.~\eqref{eq:Y_ref} implies that for $k=n$, we recover the differential operator in eq.~\eqref{eq:Lhat_q}, i.e., $\widetilde{\cL}_{n,q}=\widehat{\cL}_{n,q}$. Moreover, an easy computation shows that the recursion in eq.~\eqref{eq:u_rec} is equivalent to
\beq
\widetilde{\cL}_{k-2,q}\, \widehat{\omega}_{n+1,k} = Y_{n,k-2}\,,    \quad\text{for } k=2,\hdots,n+1\,.
\label{inhomperiods}
\eeq
This implies that we can obtain the (normalised) periods as solutions to the inhomogeneous differential equation \eqref{inhomperiods}. By direct computation, we see that the most general solution is given by ($k=2,\hdots,n+1$)
\beq
\widehat{\omega}_{n+1,k}(q) = I(Y_{n,0},Y_{n,1},\ldots,Y_{n,k-2};q) + \sum_{i = 0}^{k-3}a_i\, I(Y_{n,0},Y_{n,1},\ldots,Y_{n,i};q)     \,,
\eeq
where $a_i$ are integration constants, and we introduced the iterated integrals (cf.~eq.~\eqref{eq:II_def}):
\beq
I(f_1,\ldots,f_n;q)   = \int_{\vec{1}_{0}}^{q}\frac{\rd q'}{q'}\,f_1(q')\,I(f_2,\ldots,f_n;q') \,.
\eeq
Note the appearance of the $1/q'$ factor in the integrand, which was not present in the definition given in eq.~\eqref{eq:II_def}. This factor is included to account for the change of variables from $t$ to $q = e^{2\pi i t}$.
It is easy to check that 
\beq
I(Y_{n,0},Y_{n,1},\ldots,Y_{n,k-1};q) = \frac{1}{k!}\,\log^{k}q + \ord(q)\,.
\eeq
Our normalisation of the $\widehat{\omega}_{n,k}$ in eq.~\eqref{eq:omega_hat_norm} then implies that we must have
\beq\label{eq:iter_per}
\widehat{\omega}_{n+1,k} = I(Y_{n,0},Y_{n,1},\ldots,Y_{n,k-2};q) = I(1,Y_{n,1},\ldots,Y_{n,k-2};q)\,, \quad\text{for } k=2,\hdots,n+1.
\eeq
We see that we can write the periods as iterated integrals involving the $Y$-invariants. These iterated integrals are similar to those that have appeared in the context of the four-loop banana integral in ref.~\cite{Pogel:2022ken}. We have explicitly through $n=5$:
\begin{align}
\nonumber\widehat{\omega}_{2,2}(q) &\,= I(1;q)\,,\\
\nonumber\widehat{\omega}_{3,2}(q) &\,= I(1;q)\,, \quad \widehat{\omega}_{3,3}(q) = I(1,1;q)\\
\widehat{\omega}_{4,2}(q) &\,= I(1;q)\,, \quad \widehat{\omega}_{4,3}(q) = I(1,Y_{3,1};q)\,, \quad \widehat{\omega}_{4,4}(q) = I(1,Y_{3,1},1;q)\,,\\
\nonumber\widehat{\omega}_{5,2}(q) &\,= I(1;q)\,, \quad \widehat{\omega}_{5,3}(q) = I(1,Y_{4,1};q)\,, \quad \widehat{\omega}_{5,4}(q) = I(1,Y_{4,1},Y_{4,1};q)\,,\\ 
\nonumber&\phantom{\,= I(1;q)\,, \quad}\,\,\, \widehat{\omega}_{5,5}(q) = I(1,Y_{4,1},Y_{4,1},1;q)\,,\\
\nonumber\widehat{\omega}_{6,2}(q) &\,= I(1;q)\,, \quad \widehat{\omega}_{6,3}(q) = I(1,Y_{5,1};q)\,, \quad \widehat{\omega}_{6,4}(q) = I(1,Y_{5,1},Y_{5,2};q)\,,\\
\nonumber &\phantom{\,= I(1;q)\,, \quad}\,\,\, \widehat{\omega}_{6,5}(q) = I(1,Y_{5,1},Y_{5,2},Y_{5,1};q)\,,\quad \widehat{\omega}_{6,6}(q) = I(1,Y_{5,1},Y_{5,2},Y_{5,1},1;q)\,.
\end{align}

It is well known (cf., e.g., ref.~\cite{ChenSymbol}) that iterated integrals form a shuffle algebra, i.e., we have
\beq
I(f_1,\ldots, f_m;q) I(f_{m+1},\ldots, f_n;q)  = \sum_{\sigma\in{\Sigma(m,n)}}I(f_{\sigma_1},\ldots,f_{\sigma_n};q)\,,
\eeq
where $\Sigma(m,n)$ is the set of all shuffles of $m$ and $n-m$ elements, i.e., the subset of the permutations of $(1,\ldots,n)$ that leave the relative order of $(1,\ldots,m)$ and $(m+1,\ldots,n)$ invariant. In appendix~\ref{app:shuffle} we recall some basic facts about shuffle algebras, and we show that the shuffle product on iterated integrals, together with the reflection identity in eq.~\eqref{eq:Y_ref}, implies the following relations among the normalised periods:
\beq\label{eq:griffith}
\widehat{\underline{\omega}}_{n+1}^T\,{\bf\Sigma}_{n+1}\,\theta_q^m\widehat{\underline{\omega}}_{n+1} = \left\{\begin{array}{ll}
0\,, & { m< n}\,,\\
\prod_{k=1}^{n-2}Y_{n,k}\,, & { m= n}\,,
\end{array}\right.
\eeq
where we defined $\widehat{\underline{\omega}}_{n+1}=(\widehat{\underline{\omega}}_{n+1,1},\hdots,\widehat{\underline{\omega}}_{n+1,n+1})^T$. Note that the product on the right-hand side can be expressed through eq.~\eqref{relationYinva}. It is easy to check that these relations are equivalent to the well-known quadratic relations between periods that follow from Griffiths transversality (cf.~eq.~\eqref{quadrelations}). In other words, when the periods are expressed as iterated integrals, the quadratic relations from Griffiths transversality become simple shuffle identities among iterated integrals! It is then possible to pick a basis for the shuffle algebra (e.g., in terms of Lyndon words), and to eliminate in this way interrelations among periods due to Griffiths transversality.

\subsubsection{Extensions of Calabi-Yau operators}
\label{sec:extensions}

In the previous section we have seen that we can write the periods, which are solutions to the homogeneous Picard-Fuchs equation $\cL_{n}\Pi(x)=0$, as iterated integrals involving $Y$-invariants.
In this section we consider extensions of Calabi-Yau Picard-Fuchs differential equations of the form
\beq\label{eq:general_extension}
\cL_{n}\Pi(x) = F(x)\,,
\eeq
where $F$ is some function. Since eq.~\eqref{eq:general_extension} is an inhomogeneous linear differential equation, its general solution is obtained by adding a particular solution to the general homogeneous solution. A basis for the homogeneous solutions are the periods. Hence, it is enough to find a particular solution. This is in general a monumental task. Here we argue that the special form of Calabi-Yau operators makes it possible to write down a particular solution in a systematic manner.

We start by changing variables to the canonical coordinate $q$ on the moduli space. Equation~\eqref{eq:general_extension} then becomes
\beq\label{eq:general_extension_q}
\widehat\cL_{n,q}\,\widehat{\omega}(q) = \hat{f}(q)\,,
\eeq
where $\widehat\cL_{n,q}$ is given in eq.~\eqref{eq:Lhat_q}, and moreover we defined:
\beq
\widehat{\omega}(q) = \frac{\Pi(x(q))}{\widehat{\Pi}_{n+1,1}(q)} \textrm{~~~and~~~} \hat{f}(q) = \frac{\widehat\alpha_1(q)}{C\,\beta(x(q))}\,F(x(q))\,.
\eeq
It is easy to see that a particular solution to eq.~\eqref{eq:general_extension_q} is given by
\beq\label{eq:generic_inhom_q}
\widehat{\omega}_{\textrm{part.}}(q) = I(1,Y_{n,1},Y_{n,2},\ldots,Y_{n,2},Y_{n,1},1,\hat{f};t)\,.
\eeq

Equation~\eqref{eq:generic_inhom_q} suggest that it should be possible to write the solution for the ice cone and banana integrals in eqs.~\eqref{eq:I+l1_result} and~\eqref{eq:ban_result} in terms of the iterated integrals that represent the periods and the inhomogeneous solution, cf.~eqs.~\eqref{eq:iter_per} and~\eqref{eq:generic_inhom_q}. In the following we show that this is indeed the case. 

As a starting point, we mention the following identity:
\beq\bsp\label{eq:x_q_rel}
&\sum_{k=1}^{n+1}(-1)^{n-k+1}{\Pi}_{n+1,n-k+2}(x)\int_{\vec{1}_0}^x\,\rd x'\,G(x')\,{\Pi}_{n+1,k}(x') \\
&\qquad =(-1)^n\,{\widehat\Pi}_{n+1,1}(q)\, I(1,Y_{n,1},Y_{n,2},\ldots,Y_{n,2},Y_{n,1},1,\hat{g};q)\,, 
\esp\eeq
with
\beq
\hat{g}(q) = x(q)\widehat{\alpha}_1(q)\,\widehat\Pi_{n+1,1}(q)G(x(q))\,.
\eeq
A proof of eq.~\eqref{eq:x_q_rel} is given in appendix~\ref{app:shuffle}.

Next, we combine eq.~\eqref{eq:x_q_rel} with our results for the banana and ice cone integrals from section~\ref{subsec:mastersiterated}. We start with the banana integral given in eq.~\eqref{eq:ban_result}:
\beq\bsp\label{eq:ban_q}
        \textrm{Ban}_{l-1,1}(x;2)   &=  l! \sum_{k=1}^{l-1}\Bigg[(-1)^{l-k+1} \frac{\lambda^{(k)}_0}{(k+1)!}\,\widehat\Pi_{l-1,1}(q)\,I(1,Y_{l-2,1},\ldots,Y_{l-2,l-k-2};q)     \\
                                    &\qquad +(-1)^{l-k+1}\Pi_{l-1,l-k}(x) I(f_{l-1,k};x) \Bigg]        \\
                                    &=  (-1)^l l!\,\widehat\Pi_{l-1,1}(q) \Bigg[  \sum_{k=1}^{l-1}(-1)^{k+1} \frac{\lambda^{(k)}_0}{(k+1)!}\,I(1,Y_{l-2,1},\ldots,Y_{l-2,l-k-2};q)                 \\
                                    &\qquad +\,I(1,Y_{l-2,1},\ldots,Y_{l-2,1},1,\hat{g}_{\text{ban},l-1};q) \Bigg] \, ,
\esp\eeq
where we defined
\beq
\hat{g}_{\text{ban},l-1}(q) =  \widehat{\alpha}_1(q) \widehat{\Pi}_{l-1,1}(q)/x(q)                 \,.
\eeq
Let us give here some explicit results:
\beq\bsp\label{eq:ban_qexam}
        \hat{g}_{\text{ban},2}(q) &=  1-1q-5q^2-1q^3+11q^4 + \mathcal O(q^5)   \,,      \\
        \hat{g}_{\text{ban},3}(q) &=  1-2q-14q^2-38q^3-142q^4  + \mathcal O(q^5)   \,,     \\
        \hat{g}_{\text{ban},4}(q) &=  1-3q-27q^2-147q^3-1467q^4  + \mathcal O(q^5)    \, .
\esp\eeq

Similarly, we have:
\beq\bsp\label{eq:icq}
\cI_{l,1}^+(x;2) &=         \sum_{k=1}^{l-1}\left[c^+_{l,l-k}\,\widehat\Pi_{l-1,1}(q)\,I(1,Y_{l-2,1},\ldots,Y_{l-2,l-k-2};q)\right.     \\
                 &\qquad    \left.+ l!\,(-1)^k\, \Pi_{l-1,l-k}(x) I(f_{l-1,k},f_{0};x)            \right]  \\
                 &=         \widehat\Pi_{l-1,1}(q)\Bigg[\sum_{k=1}^{l-1}c^+_{l,l-k}\,I(1,Y_{l-2,1},\ldots,Y_{l-2,l-k-2};q) \\
                 &\qquad        -{l!}  I(1,Y_{l-2,1},\ldots,Y_{l-2,1},1,\hat{g}_{\text{ice},l};q)                \Bigg] \,,
\esp\eeq
where we defined similarly
\beq\label{gice}
\hat{g}_{\text{ice},l}(q) =  \widehat{\alpha}_1(q) \widehat{\Pi}_{l-1,1}(q)\log(x(q))/x(q)                 \,.
\eeq
Again some explicit results are:
\beq\bsp\label{eq:ban_qexam}
        \hat{g}_{\text{ice},2}(q) &=  (1-1q-5q^2)\log q -4q+6q^2+ \mathcal O(q^3)   \,,      \\
        \hat{g}_{\text{ice},3}(q) &=  (1-2q-14q^2)\log q -6q+15q^2  + \mathcal O(q^3)   \,,     \\
        \hat{g}_{\text{ice},4}(q) &=  (1-3q-27q^2)\log q -8q +28q^2 + \mathcal O(q^3)    \, .
\esp\eeq
Notice that since eq.~\eqref{gice} contains itself a logarithm, we can write:
\beq\bsp\label{eq:ban_qexam}
        I(1,Y_{l-2,1},\ldots,Y_{l-2,1},1,\hat{g}_{\text{ice},l};q)  =   I(1,Y_{l-2,1},\ldots,Y_{l-2,1},1,\hat{g}_{\text{ban},l-2},\widehat{\alpha}_1(q);q)    \, ,
\esp\eeq
which explicitly shows that the particular solution of the $l$-loop ice cone graph is obtained from the particular solution of the banana at $(l-1)$-loops, augmented by an additional integration. 

Equations~\eqref{eq:ban_q} and~\eqref{eq:icq} are main results of this paper. They express the banana and ice cone integrals in terms of iterated integrals in the canonical variable $q$. These iterated integrals are the natural generalisation of the iterated integrals of Eisenstein series that appear in the results for the two and three-loop banana integrals in $d=2$ dimensions~\cite{Bloch:2014qca,Bloch:2016izu,Adams:2017ejb,Broedel:2018iwv,Broedel:2019kmn}. It is remarkable that the letters of the iterated integrals are entirely constructed out of geometrical invariants for the Calabi-Yau $(l-1)$-folds attached to the $(l-1)$-loop banana integral, namely the period $\widehat\Pi_{l-1,1}(q)$ that is holomorphic at the MUM point $q=0$, the mirror map $x(q)$ and the $Y$-invariants $Y_{l-2,m}(q)$. We also emphasise the following point: It is easy to see that all the iterated integrals appearing in eqs.~\eqref{eq:ban_q} and~\eqref{eq:icq} diverge logarithmically at the MUM point:
\beq
I(f_1,\ldots,f_m;q) = \frac{1}{m!}\log^mq + \ord(q)\,.
\eeq
It is thus natural to assign transcendental weight $m$ to these iterated integrals. This means that the $(l-1)$-loop banana integral in eq.~\eqref{eq:ban_q} has transcendental weight $l-1$ and the $l$-loop ice cone integral in eq.~\eqref{eq:icq} has weight $l$. More precisely, they are \emph{pure functions of weight} $l-1$ and $l$~\cite{ArkaniHamed:2010gh,Broedel:2018qkq}. Since the coefficients $\lambda^{(k)}_0$ and $c^+_{l,k}$  have uniform weight $k$ and $l-k$, we see that, after normalising eqs.~\eqref{eq:ban_q} and~\eqref{eq:icq}  by the holomorphic period $\widehat\Pi_{l-1,1}(q)$ (which has transcendental weight 0 and computes a maximal cut of the integral), they are pure functions of uniform weight. This generalises the property observed for the two- and three-loop banana integrals in refs.~\cite{Broedel:2018iwv,Broedel:2019kmn} to an arbitrary number of loops.

%% file: conclusions.tex

\section{Conclusion}
\label{sec:conclusion}  

In this paper we have presented for the first time analytic results for the ice cone graphs with equal propagator masses. By analysing their generalised leading singularities in $d=2$ dimensions, we could identify a set of master integrals such that, at any number of loops $l$, the Gauss-Manin system takes a particularly convenient form. Remarkably, in this basis the homogeneous equations in the top sector organise into two copies of the periods of the Calabi-Yau varieties associated with the $(l-1)$-fold banana integrals. As a consequence, by applying the strategy developed in ref.~\cite{Bonisch:2021yfw} in the context of equal-mass banana integrals, we can solve the Gauss-Manin system for the ice cone graphs in terms of the same class of iterated integrals. The initial condition is then obtained by studying the analytic structure of the ice cone integrals, and we could identify a generating function for the integration constants. While both the basis of master integrals and the form of the generating function remains conjectural, we have tested the basis of master integrals through five loops and the generating function up to seven loops using public available tools, and we are confident that our results are valid to any number of loops. 

Our representation for the solutions involves an integration over the coordinate $x$ on the moduli space of the family of Calabi-Yau varieties. This choice of coordinates is by no means canonical. In a second part of this paper we have therefore studied how we can express the iterated integrals in terms of the canonical coordinate $q$ on the moduli space defined by the mirror map. We find that, both for the banana and ice cone integrals, the results can be written in terms of iterated integrals whose kernels involve the geometric invariants one can attach to a one-parameter family of Calabi-Yau $n$-folds. These are its mirror map $x(q)$, the period $\widehat\Pi_{n+1,1}(q)$ that is holomorphic at the MUM point and, for $n\ge 3$, the $Y$-invariants $Y_{n,m}(q)$. This is similar to the representation of the four- and five-loop banana integrals (which correspond to $n=3,4$) presented in refs.~\cite{Pogel:2022ken,xing_private}, but we observe that one needs to include a second independent $Y$-invariant starting for $n\ge 5$, consistent with the known literature on Calabi-Yau operators~\cite{bogner2013algebraic,Bognerthesis}. As a byproduct, we obtain a novel representation of the logarithmically-divergent periods as iterated integrals involving the $Y$-invariants, and we observe that in this representation the well-known quadratic relations among periods from Griffith transversality reduce to simple shuffle relations among the iterated integrals. To the best of our knowledge, this representation of the periods is new and has not yet been considered in the literature on Calabi-Yau varieties.

Our results also open new possibilities for further research. First, it would be interesting to see if one can extend the results of ref.~\cite{Pogel:2022ken} on canonical $\epsilon$-forms beyond $n=3$ and/or to ice cone integrals. An interesting direction to find an $\epsilon$-form in this context could be the extension of the algorithm of ref.~\cite{Dlapa:2022wdu} to higher-dimensional Calabi-Yau varieties. This would allow us to obtain analytic results for the higher orders in the dimensional regulator $\epsilon$ for ice cone integrals. Second, it would be interesting if one could interpret our generating functional for the initial conditions in eq.~\eqref{genseries} as a generalised $\widehat{\Gamma}$-class, similar to the conjecture of ref.~\cite{Bonisch:2021yfw} for equal-mass banana integrals. This could then possibly open the way to prove our conjectured eq.~\eqref{genseries} using the techniques of ref.~\cite{Iritani:2020qyh}. Finally, it would be interesting if one can extend our results on the iterated integrals in $q$, as well as the results of ref.~\cite{Pogel:2022ken} to families of Calabi-Yau varieties depending on more than one modulus. For example, we expect that from eq.~\eqref{eq:flat} it should be possible to obtain a representation of the periods as iterated integrals also in the multi-parameter case. We leave these topics for future work.

%% file: appendices.tex

\section{Derivation of the master integrals $I_{0,0}$ and $I_0$}
\label{app:derivationmaster}

We first write down some useful integrals. These identities are valid for suitable ranges of parameters. They are:
\begin{equation}
	\int_0^\infty (x_1 \hdots x_{l-1})^a(1+x_1+\hdots+x_{l-1})^{-b}~\mathrm dx_1\hdots dx_{l-1}	=	\frac{\Gamma(1+a)^{l-1}\Gamma(b-(l-1)(1+a))}{\Gamma(b)}\,,
\label{ir1}
\end{equation}
\begin{equation}
\begin{aligned}
	&\int_0^\infty (1+x_1)^a(x_2 \hdots x_l)^b(1+x_1+\hdots+x_l)^cx_1^d~\mathrm dx_1\hdots dx_{l-1}	=	\\
		&\hspace{3cm}	\frac{\Gamma(1+b)^{l-1}\Gamma(1+d)\Gamma(-(l-1)(b+1)+c)\Gamma(-l-a-(l-1)b-c-d)}{\Gamma(-(l-1)(1+b)+a+c)\Gamma(-c)}\,,
\label{ir2}
\end{aligned}
\end{equation}

\begin{equation}
\begin{aligned}
	&\int_0^\infty (1+x_1)^a(x_2 \hdots x_l)^b((1+x_1)(1+x_1+\hdots+x_l)-p^2x_1)^cx_1^d~\mathrm dx_1\hdots dx_{l-1}	=	\\
		&\hspace{1cm}	\frac{\Gamma(1+b)^{l-1}\Gamma(1+d)\Gamma(1+b-c-l-lb)\Gamma(-a+b-2c-d-l-lb)}{\Gamma(1-a+b-2c-l-lb)\Gamma(-c)}	\\
		&\hspace{1cm} \times \, _3F_2\Bigg(	 1+d,1-l-(l-1)b-c,-l-a-(l-1)b-2c-d ;    \\
        &\hspace{2.3cm}    \frac12-\frac l2-\frac a2-(l-1)\frac b2-c,1-\frac l2-(l-1)\frac b2-c  ; \frac {p^2}4	\Bigg)	\, ,
\label{ir3}
\end{aligned}
\end{equation}

\noindent
where the last can be derived using the second and the Gauss-hypergeometric function $_2F_1$.

Using these three integral identities we can derive the following expressions for the subtopologies. The first subtopology is the tadpole with squared propagators
\begin{equation}
	I_{0,0}	=	\int \frac{\mathrm d^dk_1}{i\pi^{d/2}} \hdots \frac{\mathrm d^dk_l}{i\pi^{d/2}} \frac1{(k_1^2-1)^2}\hdots \frac1{(k_l^2-1)^2}	=	\Gamma(2-d/2)^l	\, ,
\label{tadpoles}
\end{equation}
which is finite in $d=2$ dimensions as can easily be seen.
Next we need the subtopology which is a one-loop bubble times $(l-1)$ tadpoles. To get a finite expression, we again square the propagators belonging to the $(l-1)$ tadpoles. We find, with $p$ the momenta flowing into the graph,
\begin{equation}
\begin{aligned}
	I_0	&=	\int \frac{\mathrm d^dk_1}{i\pi^{d/2}} \hdots \frac{\mathrm d^dk_l}{i\pi^{d/2}} \frac1{(p-k_1)^2-1}\frac1{k_1^2-1}\frac1{(k_2^2-1)^2}\hdots \frac1{(k_l^2-1)^2} \\
        &=	\quad \Gamma(2-d/2)^l ~_2F_1(1,2-d/2,3/2;p^2/4)	\, .
\label{tadpoles}
\end{aligned}
\end{equation}
The last subtopology is given by the $l$-loop banana at zero ingoing momentum. For high loop orders there is no analytic form known neither for propagator powers equal to unity nor other non-negative values. But all of them are finite constants at $p^2=0$. It turns out that for the leading contribution in $d=2$ of the ice cone graphs, the banana subtopologies are not needed, as explained in section \ref{sec:lloop}.

%% file: shuffle.tex

\section{Quadratic relations from shuffle algebras}
\label{app:shuffle}
In this section we give the proofs of eqs.~\eqref{eq:griffith} and~\eqref{eq:x_q_rel}. We show that eq.~\eqref{eq:griffith} follows only from the shuffle algebra structure of iterated integrals, combined with the reflection identity of the $Y$-invariants in eq.~\eqref{eq:Y_ref}. We start with an extra section on basic facts about shuffle algebras.

\subsection{The shuffle Hopf algebra}
\label{app:shuffle_recap}

We consider a set of \emph{letters} $a_i$, from which we can form \emph{words} by concatenating letters. The \emph{length} of the word $w$, denoted by $|w|$, is the number of letters it is composed of. The \emph{empty word}, i.e., the unique word of length 0, is denoted by $1$. 

The shuffle algebra generated by these letters is the algebra $\mathbb{A}$ consisting of all $\mathbb{Q}$-linear combinations of words, together with the shuffle product:
\beq
(a_1\cdots a_k)\shuffle(a_{k+1}\cdots a_\ell)=\sum_{\sigma\in\Sigma(k,\ell)}a_{\sigma_1}\cdots a_{\sigma_\ell}\,,
\eeq
where $\Sigma(k,\ell)$ is the set of all shuffles of $k$ and $\ell-k$ elements, i.e., the subset of the permutations of $(1,\ldots,\ell)$ that leave the relative order of $(1,\ldots,k)$ and $(k+1,\ldots,\ell)$ invariant. The shuffle algebra is graded by the length of the words,
\beq
\mathbb{A}=\bigoplus_{\ell=0}^\infty\mathbb{A}_\ell\,,
\eeq
where $\mathbb{A}_\ell$ is the vector space generated by all words of length $\ell$.

Every shuffle algebra is a Hopf algebra, whose coproduct is the deconcatenation of words and the antipode is the reversal of words (up to a sign):
\beq\bsp
\Delta(a_1\cdots a_\ell) &\,= \sum_{k=0}^{\ell+1}(a_1\cdots a_{k})\otimes(a_{k+1}\cdots a_\ell)\,,\\
S(a_1\cdots a_\ell) &\,= (-1)^\ell\,a_\ell\cdots a_1 \,.
\esp\eeq
The counit is the projection onto words of length zero, i.e., if $w$ is a word:
\beq
\varepsilon(w) = \left\{\begin{array}{ll}
0\,,& |w|>0\,,\\
1\,,& |w|=0\,.\end{array}\right.
\eeq

It is one of the defining axioms of a Hopf algebra that the multiplication $m$, the coproduct $\Delta$, the antipode $S$ and the counit $\varepsilon$ are related by
\beq
m(\textrm{id}\otimes S)\Delta = \varepsilon\,.
\eeq
This implies that the following identity holds in every shuffle algebra ($\ell>0$):
\beq
\sum_{k=0}^{\ell+1}(a_1\cdots a_{k})\shuffle S(a_{k+1}\cdots a_{\ell}) =\sum_{k=0}^{\ell+1}(-1)^{\ell-k}(a_1\cdots a_{k})\shuffle(a_{\ell}\cdots a_{k+1}) = 0\,.
\eeq
Since iterated integrals form a shuffle algebra, this implies that we have the following relation among iterated integrals:
\beq\bsp\label{eq:shuffle_antipode_relation}
\sum_{k=0}^{\ell+1}&I(f_1,\ldots, f_{k};q)\, I^S(f_{k+1},\ldots, f_{\ell};q)\\
&\, =\sum_{k=0}^{\ell+1}(-1)^{\ell-k}I(f_1,\ldots, f_{k};q)\,I(f_{\ell},\ldots, f_{k+1};q) = 0\,,
\esp\eeq
where we defined
\beq
I^S(f_1,\ldots,f_m;q) = (-1)^m\,I(f_m,\ldots,f_1;q)\,.
\eeq

\subsection{Proof of eq. \eqref{eq:griffith}}

In this appendix we prove eq.~\eqref{eq:griffith}. While eq.~\eqref{eq:griffith} is equivalent to the well-known quadratic relations among periods following from Griffiths transversality, we show here that it is possible to prove eq.~\eqref{eq:griffith} only using the shuffle algebra structure of iterated integrals, without using Griffiths transversality as an input.

We start by defining, for $m\ge0$,
\begin{align}
\underline\Omega_{n}^{[m]} &\, = \Big(\underbrace{0,\ldots,0}_{m},1,I(Y_{n,m};q),\ldots,I(Y_{n,m},\ldots,Y_{n,n-1};q)\Big)^T\,.
\end{align}
Note that for $m=0,1,2$ we have the relations:
\beq
\underline{\widehat{\omega}}_{n+1} = \underline\Omega_{n}^{[0]},    \quad
\theta_q\,\underline{\widehat{\omega}}_{n+1} = \underline\Omega_{n}^{[1]} \quad\text{and}\quad
\theta_q^2\,\underline{\widehat{\omega}}_{n+1} = Y_{n,1}\,\underline\Omega_{n}^{[2]} \,.
\eeq
For $m>2$, the relation between $\underline\Omega_{n}^{[m]}$ and the derivatives of $\underline{\widehat{\omega}}_n$ is more complicated, and involves linear combinations:
\beq\label{eq:om_to_Om}
\theta_q^m\underline{\widehat{\omega}}_{n+1}=\left(\prod_{p=0}^{m-1}Y_{n,p}\right)\, \underline\Omega_{n}^{[m]} + \sum_{r=1}^{m-1}C_r\,\underline\Omega_{n}^{[r]}\,,
\eeq
where $C_r$ are sums of products of $Y$-invariants and their derivatives. Their explicit form is irrelevant here. The important point is that it is easy to see from eq.~\eqref{eq:om_to_Om} that eq.~\eqref{eq:griffith} is equivalent to
\beq\label{eq:Griff_Om}
{\underline{\Omega}}_{n}^{[0]T}\,{\bf\Sigma}_{n+1}\,{\underline{\Omega}}_{n}^{[m]} = \delta_{mn}\,.
\eeq
This is manifest for $m=n$. For $m<n$, we have
\beq\bsp
{\underline{\Omega}}_{n}^{[0]T}\,{\bf\Sigma}_{n+1}\,{\underline{\Omega}}_{n}^{[m]}
&\,=\sum_{k=1}^{n+1}(-1)^{n-k+1}{{\Omega}}_{n,n-k+2}^{[0]}\,{{\Omega}}_{n,k}^{[m]}\\
&\,=(-1)^{n-m}{{\Omega}}_{n,n-m+1}^{[0]}+{{\Omega}}_{n,n+1}^{[m]}+\sum_{k=m+2}^{n}(-1)^{n-k+1}{{\Omega}}_{n,n+2-k}^{[0]}\,{{\Omega}}_{n,k}^{[m]}\\
&\,=(-1)^{n-m}I(Y_{n,0},\ldots,Y_{n,n-m-1};q)+I(Y_{n,m},\ldots,Y_{n,n-1};q)+\\
&\,\qquad+\sum_{k=m+2}^{n}(-1)^{n-k+1}I(Y_{n,0},\ldots,Y_{n,n-k};q)\,I(Y_{n,m},\ldots,Y_{n,k-2};q)\\
&\,=(-1)^{n-m}I(Y_{n,n-1},\ldots,Y_{n,m};q)+I(Y_{n,m},\ldots,Y_{n,n-1};q)+\\
&\,\qquad+\sum_{k=m+2}^{n}(-1)^{n-k+1}I(Y_{n,n-1},\ldots,Y_{n,k-1};q)\,I(Y_{n,m},\ldots,Y_{n,k-2};q)\\
&\,=I^S(Y_{n,m},\ldots,Y_{n,n-1};q)+I(Y_{n,m},\ldots,Y_{n,n-1};q)+\\
&\,\qquad+\sum_{k=m+2}^{n}I(Y_{n,m},\ldots,Y_{n,k-2};q)\,I^S(Y_{n,k-1},\ldots,Y_{n,n-1};q)\\
&\,=0\,,
\esp\eeq
where in the fourth equality we have used the reflection identity~\eqref{eq:Y_ref} for the $Y$-invariants, and the last step follows immediately from eq.~\eqref{eq:shuffle_antipode_relation}.

\subsection{Proof of eq. \eqref{eq:x_q_rel}}
 
We use the notations and conventions of section~\ref{sec:extensions}. The proof follows exactly the same argument as for the proof of eq.~\eqref{eq:Griff_Om}. We have:
\begin{align}
&\underline{\Pi}_{n+1}(x)^T{\bf\Sigma}_{n+1}\int_{\vec{1}_0}^x\,\rd x'\,G(x')\,\underline{\Pi}_{n+1}(x')  \nonumber \\
& =   \sum_{k=1}^{n+1}(-1)^{n-k+1}{\Pi}_{n+1,n-k+2}(x)\int_{\vec{1}_0}^x\,\rd x'\,G(x')\,{\Pi}_{n+1,k}(x')\nonumber \\
& =   {\widehat\Pi}_{n+1,1}(q)\sum_{k=1}^{n+1}(-1)^{n-k+1}\widehat{\omega}_{n+1,n-k+2}(q)\int_{\vec{1}_0}^q\,\frac{\rd q'}{q'}\,\hat{g}(q')\,\widehat{\omega}_{n+1,k}(q')\nonumber\\
& =   {\widehat\Pi}_{n+1,1}(q)  \sum_{k=1}^{n+1}(-1)^{n-k+1}I(Y_{n,0},Y_{n,1},\ldots,Y_{n,n-k};q)    \nonumber\int_{\vec{1}_0}^q\frac{\rd q'}{q'}\hat{g}(q')I(Y_{n,0},Y_{n,1},\ldots,Y_{n,k-2};q')\nonumber\\
& =   {\widehat\Pi}_{n+1,1}(q) \sum_{k=1}^{n+1}(-1)^{n-k+1}I(Y_{n,0},Y_{n,1},\ldots,Y_{n,n-k};q)\,I(\hat{g},Y_{n,0},Y_{n,1},\ldots,Y_{n,k-2};q)\nonumber\\
& =   {\widehat\Pi}_{n+1,1}(q)  \sum_{k=1}^{n+1}(-1)^{n-k+1}I(Y_{n,n-1},Y_{n,n-2},\ldots,Y_{n,k-1};q)\,I(\hat{g},Y_{n,0},Y_{n,1},\ldots,Y_{n,k-2};q)\nonumber\\
& =   {\widehat\Pi}_{n+1,1}(q)  \sum_{k=1}^{n+1}I(\hat{g},Y_{n,0},Y_{n,1},\ldots,Y_{n,k-2};q)I^S(Y_{n,k-1},\ldots,Y_{n,n-2},Y_{n,n-1};q)\nonumber\\[1ex]
& =   -{\widehat\Pi}_{n+1,1}(q)  I^S(\hat{g},Y_{n,0},Y_{n,1},\ldots,Y_{n,n-1};q)\nonumber\\[1ex]
& =   (-1)^n{\widehat\Pi}_{n+1,1}(q)  I(Y_{n,n-1},\ldots,Y_{n,1},Y_{n,0},\hat{g};q) \nonumber\\[1ex]
& =   (-1)^n{\widehat\Pi}_{n+1,1}(q)  I(1,Y_{n,1},Y_{n,2},\ldots,Y_{n,2},Y_{n,1},1,\hat{g};q) \,,
\end{align}
where we used the reflection property of the $Y$-invariants in eq.~\eqref{eq:Y_ref} and the last step follows immediately from eq.~\eqref{eq:shuffle_antipode_relation}.